\documentclass[12pt,english,floatfix,nofootinbib,superscriptaddress,aps,prd,preprint]{revtex4}
\usepackage[utf8]{inputenc}
\usepackage{float}
\usepackage{array}
\usepackage{bbold}
\usepackage{lipsum}
\usepackage{dsfont}
\usepackage{graphicx}
\usepackage{amsmath,amsthm,amsfonts,amssymb}
\usepackage{graphicx}
\usepackage[english]{babel} 
\usepackage{color}
\usepackage{tensor}
\usepackage{esint}
\usepackage[dvips]{epsfig}
\usepackage[dvips]{graphicx}
\usepackage{float}
\usepackage{units}
\usepackage{textcomp}
\usepackage{mathrsfs}
\usepackage{amsmath}
\usepackage[makeroom]{cancel}
\usepackage{amssymb}
\usepackage{amsbsy}
\usepackage{amsfonts}
\usepackage{amssymb,mathrsfs,xcolor}
\usepackage{esint}
\usepackage{braket}
\usepackage{array}
\usepackage{graphicx}

\usepackage{wasysym}
\usepackage{multirow}
\usepackage{wrapfig}
\usepackage{subfig}

\usepackage{stmaryrd}
\usepackage{upgreek}

\makeatletter

\makeatletter\usepackage{babel}

\usepackage{hyperref}
\hypersetup{
    colorlinks,
    citecolor=blue,
    filecolor=green,
    linkcolor=purple,
    urlcolor=red,
}

\usepackage{slashed}

\newcommand{\ie}{\begin{equation}}
\newcommand{\fe}{\end{equation}}
\newcommand{\se}{\begin{eqnarray}}
\newcommand{\ff}{\end{eqnarray}}

\begin{document}

\title{Particle creation and evaporation in Kalb--Ramond gravity}

\author{A. A. Ara\'{u}jo Filho}
\email{dilto@fisica.ufc.br}

\affiliation{Departamento de Física, Universidade Federal da Paraíba, Caixa Postal 5008, 58051-970, João Pessoa, Paraíba,  Brazil.}


\date{\today}

\begin{abstract}

In this work, we examine particle creation and the evaporation process in the context of Kalb--Ramond gravity. Specifically, we build upon two existing solutions from the literature \cite{yang2023static}  (Model I) and \cite{Liu:2024oas} (Model II), both addressing a static, spherically symmetric configuration. For this study, we focus on the scenario in which the cosmological constant vanishes. The analysis begins by examining bosonic particles to investigate Hawking radiation. Using the Klein–Gordon equation, the Bogoliubov coefficients are derived, highlighting the role of the parameter $\ell$, which governs Lorentz symmetry breaking, in introducing corrections to the amplitude of particle production. This forms the basis for calculating the Hawking temperature. The study further explores Hawking radiation through the tunneling mechanism, where divergent integrals are solved using the residue method. The particle creation density is also computed for fermionic particle modes. Additionally, greybody bounds are evaluated for bosons and fermions as well. Finally, we analyze the deviation of our results from those predicted by general relativity. In a general panorama, Model I exhibits the highest particle creation densities and the fastest evaporation process, whereas Model II shows the largest greybody factor intensities.

\end{abstract}

\maketitle
     
\tableofcontents

\pagebreak

\section{Introduction}

Lorentz symmetry ensures that physical laws are invariant under transformations between inertial reference frames. Although confirmed through numerous experiments, some theoretical models suggest possible violations in extreme energy regimes. Such deviations have been investigated in different approaches \cite{4,araujo2023thermodynamics,heidari2023gravitational,2,1,3,6,5,8,7}. Lorentz symmetry breaking (LSB) generally occurs in two ways: explicit and spontaneous \cite{bluhm2006overview}. In explicit LSB, the governing equations themselves lack Lorentz invariance, leading to directional dependencies in observable quantities. Spontaneous LSB, in contrast, arises when the fundamental equations remain invariant, but the vacuum state fails to reflect this symmetry, giving rise to remarkable consequences \cite{bluhm2008spontaneous}.

The study of spontaneous LSB explores deviations from standard Lorentz invariance, often framed within the Standard Model Extension (SME) \cite{liu2024shadow,12,KhodadiPoDU2023,AraujoFilho:2024ykw,13,11,9,10,filho2023vacuum}. In particular, bumblebee models are commonly used in this context, where a vector field known as the bumblebee field develops a non--zero vacuum expectation value (VEV), introducing directional dependence in particle interactions and altering local spacetime symmetries. Such symmetry breaking phenomena have received attention for their influence on thermodynamic behaviors in gravitational settings \cite{aa2021lorentz,paperrainbow,aa2022particles,araujo2021higher,araujo2021thermodynamic,araujo2022does,aaa2021thermodynamics,anacleto2018lorentz,petrov2021bouncing2,reis2021thermal,araujo2022thermal}.

Furthermore, solutions describing static, spherically symmetric spacetimes under bumblebee gravity were first formulated in Ref. \cite{14}. Extensions of these ones to Schwarzschild--like geometries have been widely examined in various scenarios, such as gravitational lensing \cite{15}, accretion processes \cite{18,17}, quasinormal modes \cite{19,Liu:2022dcn}, and Hawking radiation \cite{kanzi2019gup}.

The investigation of black hole solutions beyond standard models has led to diverse approaches, including modified (A)dS--Schwarzschild spacetimes. It was proposed a variant where vacuum conditions were relaxed, altering the gravitational background \cite{20}. Afterwards, it was expanded on this by introducing bumblebee black holes, which incorporate a vector field with a nonzero time--like component \cite{24,23,22,21}.

On the other hand, a different route to address Lorentz symmetry breaking involves the Kalb--Ramond field, a rank--two antisymmetric tensor field \cite{42,maluf2019antisymmetric} linked to bosonic string theory \cite{43}. When this field couples nonminimally with gravity and develops a vacuum expectation value, spontaneous Lorentz symmetry breaking arises. A static, spherically symmetric black hole within this theory was examined in \cite{44}, while subsequent analyses focused on particle dynamics near such configurations \cite{45}. Additionally, rotating Kalb--Ramond black holes have been investigated for their gravitational lensing effects and shadow patterns \cite{46}.

In addition, Ref. \cite{yang2023static} presents a set of new exact solutions for static and spherically symmetric spacetimes, developed both with and without a cosmological constant. These ones are constructed within a framework where the Kalb--Ramond field has a non--zero vacuum expectation value. Later, Ref. \cite{Liu:2024oas} reveals an alternative black hole configuration that had not been previously explored in \cite{yang2023static}.

Furthermore, Hawking’s work established a crucial connection between quantum mechanics and gravity, laying the groundwork for quantum gravity theory \cite{o1,o11,o111}. His research revealed that black holes can emit thermal radiation and slowly diminish over time, a phenomenon known as \textit{Hawking} radiation \cite{eeeOvgun:2019ygw,gibbons1977cosmological,eeeKuang:2017sqa,eeeKuang:2018goo,eeeOvgun:2019jdo,eeeOvgun:2015box}. This result, derived from quantum field theory near a black hole’s event horizon, has significantly advanced the study of black hole thermodynamics and quantum processes in strong gravitational fields \cite{sedaghatnia2023thermodynamical,o9,o6,araujo2024dark,o4,o3,o7,aa2024implications,araujo2023analysis,o8}. Later, Kraus and Wilczek \cite{o10}, followed by Parikh and Wilczek \cite{o13,o12,011}, interpreted \textit{Hawking} radiation as a tunneling effect within a semi--classical framework. This approach has since been widely applied across various black hole models, offering important perspectives on black hole properties \cite{giavoni2020quantum,johnson2020hawking,senjaya2024bocharova,vanzo2011tunnelling,calmet2023quantum,mirekhtiary2024tunneling,anacleto2015quantum,mitra2007hawking,silva2013quantum,medved2002radiation,del2024tunneling,zhang2005new,touati2024quantum}.

In this manner, this study investigates particle creation and black hole evaporation within the framework of Kalb--Ramond gravity. We build on two established solutions from the literature \cite{yang2023static, Liu:2024oas}, both of which describe static, spherically symmetric spacetimes, focusing specifically on cases where the cosmological constant is zero. This work initiates its investigation of Hawking radiation by focusing on bosonic particles. Through the application of the Klein–Gordon equation, the Bogoliubov coefficients are determined, revealing the influence of the parameter $\ell$, which governs Lorentz symmetry breaking and introduces corrections to the particle production amplitude. This analysis serves as the foundation for deriving the Hawking temperature. The tunneling mechanism is subsequently examined, with divergent integrals solved using the residue method. Next, the particle creation density is also addressed to fermionic modes. Greybody bounds are also analyzed for both bosonic and fermionic particles. In addition, we analyze the departure of our results from those ones predicted by general relativity. Here, to accomplish all calculations, we adopt natural units, setting $G = c = \hbar = k_B = 1$. 


\section{The general panorama}

Recently, two distinct black hole solutions have emerged in the literature. To maintain clarity, we will refer to the first solution as Model I \cite{yang2023static}
\ie
\begin{split}
\label{model1}
\mathrm{d}s^{2}  = & - \left( \frac{1}{1-\ell} - \frac{2M}{r}   \right) \mathrm{d}t^{2} + \frac{\mathrm{d}r^{2}}{\frac{1}{1-\ell} - \frac{2M}{r} } \\
& + r^{2}\mathrm{d}\theta^{2} + r^{2} \sin^{2}\mathrm{d}\varphi^{2},
\end{split}
\fe
and the second as Model II \cite{Liu:2024oas}
\ie
\begin{split}
\label{model2}
\mathrm{d}s^{2} = & - \left( 1 - \frac{2M}{r}   \right) \mathrm{d}t^{2} + \frac{(1 - \ell)}{1 - \frac{2M}{r} } \, \mathrm{d}r^{2} + r^{2}\mathrm{d}\theta^{2} \\
& + r^{2} \sin^{2}\mathrm{d}\varphi^{2}.
\end{split}
\fe
Notably, Model I has been extensively studied in the literature, with various works addressing its quasinormal modes \cite{araujo2024exploring}, greybody factors \cite{guo2024quasinormal}, gravitational lensing properties \cite{junior2024gravitational}, spontaneous symmetry--breaking constraints \cite{junior2024spontaneous}, circular motion and quasi--periodic oscillations (QPOs) near the black hole \cite{jumaniyozov2024circular}, and accretion of Vlasov gas \cite{jiang2024accretion}. Additionally, an electrically charged extension of this model has been developed \cite{duan2024electrically}, along with analyses of its implications \cite{heidari2024impact,al-Badawi:2024pdx,aa2024antisymmetric,Zahid:2024ohn,chen2024thermal,hosseinifar2024shadows}.

For Model II, recent studies have explored entanglement degradation \cite{liu2024lorentz}. It is worth noting that this second solution bears a structural resemblance to the one encountered in bumblebee gravity under the metric formalism, differing only by the sign in \(\ell\).

To maintain clarity, the following sections will examine Model I and Model II separately, allowing for detailed calculations of particle creation for both bosons and fermions. Additionally, the evaporation process for each model will be thoroughly analyzed.


\section{Model I}

\subsection{Bosonic modes}


\subsubsection{The \textit{Hawking} radiation}

We begin by examining a general spherically symmetric spacetime
\ie
\label{mainmetric}
\mathrm{d} s^2  = -\mathfrak{f}(r)\mathrm{d}t^2 + \frac{1}{\mathfrak{g}(r)}\mathrm{d}r^2 + r^2\mathrm{d}\Omega^2,
\fe    
with 
\ie
\mathfrak{f}(r) = \mathfrak{g}(r) =  \frac{1}{1-\ell} - \frac{2M}{r} . 
\fe 
Building on this foundation, we aim to explore the influence of the Lorentz violation, represented by \(\ell\), on the emission of \textit{Hawking} radiation. In his work \cite{hawking1975particle}, Hawking analyzed the scalar field, which he formulated as 
\ie
\frac{1}{\sqrt{-\mathfrak{g}}}\partial_{\mu}(\mathfrak{g}^{\mu\nu}\sqrt{-\mathfrak{g}}\partial_{\nu}\Phi) = 0,
\fe
in the framework of curved spacetime defined by the Schwarzschild case. The field operator can be formulated as
\ie
\Phi=\sum_i \left (\mathfrak{f}_i  \mathfrak{a}_i+\bar{\mathfrak{f}}_i  \mathfrak{a}^\dagger_i \right)=\sum_i \left( \mathfrak{p}_i  \mathfrak{b}_i + \bar{\mathfrak{p}}_i  \mathfrak{b}^\dagger_i + \mathfrak{q}_i  \mathfrak{c}_i + \bar{\mathfrak{q}}_i  \mathfrak{c}^\dagger_i \right ) .
\fe
In this setting, the solutions \( \mathfrak{f}_i \) and \( \bar{\mathfrak{f}}_i \) (with \( \bar{\mathfrak{f}}_i \) as the complex conjugate) represent purely ingoing components of the wave equation. Solutions \( \mathfrak{p}_i \) and \( \bar{\mathfrak{p}}_i \) describe exclusively outgoing components, while \( \mathfrak{q}_i \) and \( \bar{\mathfrak{q}}_i \) correspond to solutions without any outgoing parts. Here, the operators \( \mathfrak{a}_i \), \( \mathfrak{b}_i \), and \( \mathfrak{c}_i \) serve as annihilation operators, with their counterparts \( \mathfrak{a}_i^\dagger \), \( \mathfrak{b}_i^\dagger \), and \( \mathfrak{c}_i^\dagger \) acting as creation operators. Our objective is to establish that the solutions \( \mathfrak{f}_i \), \( \bar{\mathfrak{f}}_i \), \( \mathfrak{p}_i \), \( \bar{\mathfrak{p}}_i \), \( \mathfrak{q}_i \), and \( \bar{\mathfrak{q}}_i \) are affected by Lorentz violation. Specifically, we aim to show how the Lorentz--violating parameter alters Hawking’s initial solutions.

Due to the spherical symmetry upheld by both classical Schwarzschild spacetime and Kalb--Ramond gravity, the solutions for incoming and outgoing waves can be expanded in terms of spherical harmonics. In the exterior region of the black hole, these wave solutions can be formulated as \cite{calmet2023quantum,heidari2024quantum}:
\begin{eqnarray}
\mathfrak{f}_{\omega^\prime l m} &=& \frac{1}{\sqrt{2 \pi \omega^\prime} r }  \mathfrak{F}_{\omega^\prime}(r) e^{i \omega^\prime \mathfrak{v}} Y_{lm}(\theta,\phi)\ , \\ 
\mathfrak{p}_{\omega l m} &=& \frac{1}{\sqrt{2 \pi \omega} r }  \mathfrak{I}_\omega(r) e^{i \omega \mathfrak{u}} Y_{lm}(\theta,\phi). 
\end{eqnarray}
In this context, \(\mathfrak{v}\) and \(\mathfrak{u}\) serve as the advanced and retarded coordinates, respectively; $\mathfrak{F}_{\omega^\prime}(r)$ and $\mathfrak{I}_\omega(r)$ account for functions that depend on $r$ coordinate. Within the classical Schwarzschild framework, they are given by
\[ \mathfrak{v} = t + r + 2 M \ln \left \lvert \frac{r}{2 M } -1 \right \rvert ,\]
and
\[ \mathfrak{u} = t - r - 2 M \ln \left \lvert \frac{r}{2 M } -1 \right \rvert. \]
Using these expressions, we aim to identify the Lorentz--violating corrections that emerge from these coordinate functions. A straightforward way to approach this analysis is to consider a particle’s motion along a geodesic within the background spacetime, parameterized by an affine parameter \(\lambda\). This allows us to describe the particle’s momentum by the expression
\ie
p_{\mu}=\mathfrak{g}_{\mu\nu}\frac{\mathrm{d}x}{\mathrm{d}\lambda}^\nu.
\fe
This momentum remains conserved along the geodesic path. Moreover, the expression
\ie
\mathfrak{L} = \mathfrak{g}_{\mu\nu} \frac{\mathrm{d}x^\mu}{\mathrm{d}\lambda} \frac{\mathrm{d}x^\nu}{\mathrm{d}\lambda}
\fe
is also conserved along geodesic paths. For particles with mass, we assign \(\mathfrak{L} = -1\) and set \(\lambda = \tau\), where \(\tau\) denotes the particle's proper time. In the case of massless particles, which are the primary focus here, we use $\mathfrak{L} = 0$. By considering a stationary, spherically symmetric metric as specified in and analyzing radial geodesics (where \(p_\varphi = L = 0\)) within the equatorial plane (\(\theta = \pi/2\)), we can derive the relevant expressions
\ie
E =  \mathfrak{f}(r) \dot{t},
\fe
where \( E = -p_{t} \) (which holds true only asymptotically), and a dot indicates differentiation with respect to \(\lambda\) (i.e., \(\mathrm{d}/\mathrm{d}\lambda\)). Following this approach, we also find
\ie
\label{eq:drdl}
    \left( \frac{\mathrm{d}r}{\mathrm{d}\lambda} \right)^2 = \frac{E^2}{\mathfrak{f}(r)\mathfrak{g}(r)^{-1}},
\fe
and with some algebraic manipulations, we get
\ie
    \frac{\mathrm{d}}{\mathrm{d}\lambda}\left(t\mp r^{*}\right) = 0,
\fe
where \( r^* \) is known as the tortoise coordinate, defined as
\ie
\mathrm{d}r^{*} = \frac{\mathrm{d}r}{\sqrt{\mathfrak{f}(r)\mathfrak{g}(r)}}.
\fe
The conserved quantities are represented by the advanced and retarded coordinates, \(\mathfrak{v}\) and \(\mathfrak{u}\), respectively, which are defined as follows:
\ie
\mathfrak{v} = t + r^{*} = t  + r(1- \ell) + 2 (l-1)^2 M \ln |2 (\ell-1) M+r|,
\fe
and
\ie
\mathfrak{u} = t - r^{*} = t - r(1- \ell) - 2 (l-1)^2 M \ln |2 (\ell-1) M+r| .
\fe

By reworking the expression for the retarded coordinate, we arrive at
\ie
\label{lagsd}
\frac{\mathrm{d}u}{\mathrm{d}\lambda}=\frac{2E}{\mathfrak{f}(r)}.
\fe
Along an ingoing geodesic parameterized by \(\lambda\), the advanced coordinate \(\mathfrak{u}\) can be expressed as a function, \(\mathfrak{u}(\lambda)\). Deriving this function involves two primary steps: first, representing \(r\) as a function of \(\lambda\), and then carrying out the integral specified in Eq. \eqref{lagsd}. The exact form of \(\mathfrak{u}(\lambda)\) ultimately impacts the resulting Bogoliubov coefficients, which play a critical role in describing the quantum emission of the black hole. Proceeding with this, we use \(\mathfrak{f}(r)\) and \(\mathfrak{g}(r)\) and integrate the square root of Eq. \eqref{eq:drdl} over the range \(\tilde{r} \in [\mathfrak{r}, r]\) (where $\mathfrak{r}$ represents the event horizon), corresponding to \(\tilde{\lambda} \in [0, \lambda]\).
In this manner, we have
\ie
\label{rfgr}
r= 2M (1 - \ell) - E\lambda.
\fe
It is worth noting that, to achieve this result, we selected the negative sign in the square root when solving Eq. \eqref{eq:drdl}, corresponding to the ingoing geodesic.

Next, we utilize \(r(\lambda)\) to carry out the integration, yielding 
\ie
u(\lambda) = 2 E \lambda -4 (1-\ell)^{2} M \ln \left( \frac{\lambda}{C} \right),
\fe
where \(C\) is a constant of integration. It is worth noting that, far from the event horizon, the expression simplifies to $u(\lambda) \approx 2 E \lambda$ \cite{parker2009quantum}. However, in its vicinity, it becomes $u(\lambda) \approx -4 (1-\ell)^2 M \ln \left( \frac{\lambda}{C} \right)$, which represents the configuration that we shall address in the subsequent calculations.

Additionally, the principles of geometric optics provide a connection between the ingoing and outgoing null coordinates. This relation is given by \(\lambda = (\mathfrak{v}_0 - \mathfrak{v})/D\), where \(\mathfrak{v}_0\) represents the advanced coordinate at the horizon reflection point (\(\lambda = 0\)), and \(D\) is a constant \cite{calmet2023quantum}.

With these preliminary steps established, we now derive the outgoing solutions to the modified Klein--Gordon equation, incorporating the Lorentz--violating term $\ell$. The resulting expressions can be formulated as follows:
\ie
\mathfrak{p}_{\omega} =\int_0^\infty \left ( \alpha_{\omega\omega^\prime} \mathfrak{f}_{\omega^\prime} + \beta_{\omega\omega^\prime} \bar{ \mathfrak{f}}_{\omega^\prime}  \right)\mathrm{d} \omega^\prime,
\fe
with \(\alpha_{\omega\omega^\prime}\) and \(\beta_{\omega\omega^\prime}\) denoting the Bogoliubov coefficients \cite{parker2009quantum,hollands2015quantum,wald1994quantum,fulling1989aspects}
\begin{equation}
\begin{split}
    \alpha_{\omega\omega^\prime}=& -i\mathfrak{K}e^{i\omega^\prime \mathfrak{v}_0}e^{\pi \left[2 M (1- \ell)^{2} \right]\omega} \int_{-\infty}^{0} \,\mathrm{d}x\,\Big(\frac{\omega^\prime}{\omega}\Big)^{1/2}e^{\omega^\prime x} e^{i\omega\left[4M(1-\ell)^{2})\right]\ln\left(\frac{|x|}{CD}\right)},
    \end{split}
\end{equation}
and
\begin{equation}
\begin{split}
    \beta_{\omega\omega'} &= i\mathfrak{K}e^{-i\omega^\prime \mathfrak{v}_0}e^{-\pi \left[2 M (1- \ell)^{2} \right]\omega}
    \int_{-\infty}^{0} \,\mathrm{d}x\,\left(\frac{\omega^\prime}{\omega}\right)^{1/2}e^{\omega^\prime x}  e^{i\omega\left[4M(1-\ell)^{2})\right]\ln\left(\frac{|x|}{CD}\right)},
    \end{split}
\end{equation}
with $\mathfrak{K}$ representing the normalization constant. This demonstrates that the quantum amplitude for particle production is impacted by Lorentz--violating corrections, represented by \(\ell\) in the metric.

Notably, even though the quantum gravitational correction influences the quantum amplitude, the power spectrum continues to exhibit blackbody characteristics at this stage. To verify this, it is sufficient to calculate
\ie
    |\alpha_{\omega\omega'}|^2 = e^{\big(8\pi M (1-\ell)^{2} \big)\omega}|\beta_{\omega\omega'}|^2\,.
\fe
Examining the flux of outgoing particles within the frequency range \(\omega\) to \(\omega + \mathrm{d}\omega\) \cite{o10}, we find:
\ie
    \mathfrak{P}(\omega, \ell)=\frac{\mathrm{d}\omega}{2\pi}\frac{1}{\left \lvert\frac{\alpha_{\omega\omega^\prime}}{\beta_{\omega\omega^\prime}}\right \rvert^2-1}\, ,
\fe
or
\ie
    \mathfrak{P}(\omega, \ell)=\frac{\mathrm{d}\omega}{2\pi}\frac{1}{e^{\left(8\pi M (1-\ell)^{2}\right)\omega}-1}\,.
\fe
Here, one important aspect is worthy to be noted: if we compare above expression with the Planck distribution
\ie
    \mathfrak{P}(\omega, \ell)=\frac{\mathrm{d}\omega}{2\pi}\frac{1}{e^{\frac{\omega}{T}}-1},
\fe
we have
\ie
\label{hawtemp1}
    T = \frac{1}{8 \pi  (1- \ell)^2 M}.
\fe

At this point, the entire analysis presented in this subsection for Model I, including the Bogoliubov coefficients and tunneling amplitude, was fundamentally based by Ref. \cite{calmet2023quantum}. As we shall see, a similar procedure will be applied in the case of Model II as well.

\begin{figure}
    \centering
      \includegraphics[scale=0.55]{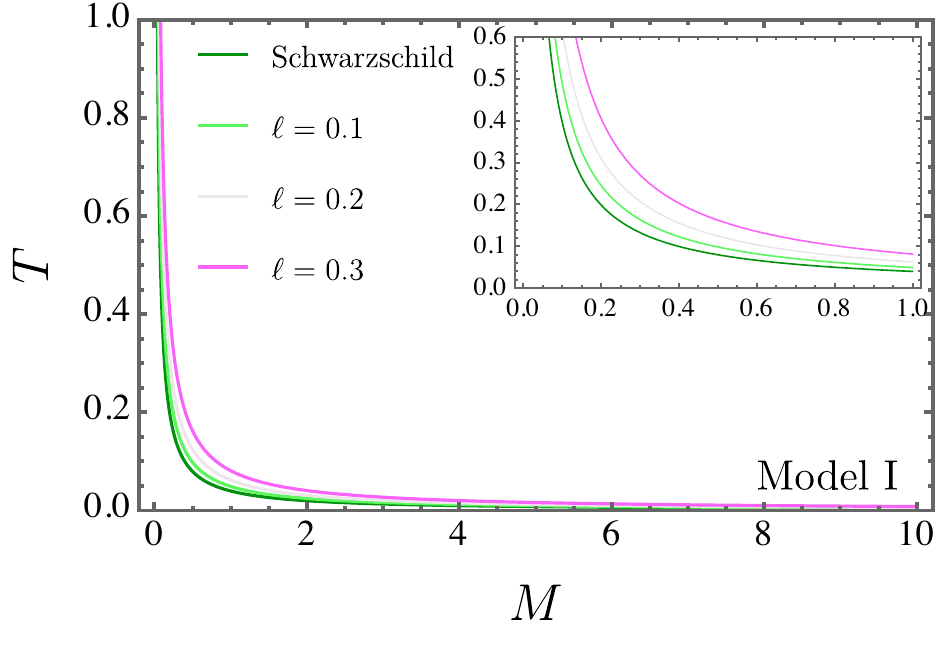}
    \caption{The \textit{Hawking} temperature $T$ as a function of mass \( M \) for various values of \( \ell \) in Model I.}
    \label{hawkingt}
\end{figure}

As we shall see in the evaporation subsection, the \textit{Hawking} temperature derived in Eq. (\ref{hawtemp}) aligns with the temperature calculated through surface gravity in Eq. (\ref{haah}), as expected. To provide a clearer understanding of this thermal property, its behavior is illustrated in Fig. \ref{hawkingt}, where a comparison with the standard Schwarzschild case is also presented. Generally, an increase in \(\ell\) results in a higher \textit{Hawking} temperature.

In other words, this implies that a black hole described by a Lorentz--violating metric radiates similarly to a \textit{greybody}, with an effective temperature \(T\) defined by \eqref{hawtemp1}.

It is essential to point out that energy conservation for the system as a whole has not been fully considered thus far. With each emission of radiation, the black hole’s total mass decreases, leading to a gradual contraction. In the following section, we will apply the tunneling method introduced by Parikh and Wilczek \cite{011} to incorporate this effect.


\subsubsection{The tunneling process}

To incorporate energy conservation in calculating the radiation spectrum of our black hole solution, we employ the method presented in \cite{011, vanzo2011tunnelling, parikh2004energy, calmet2023quantum}. Adopting the Painlevé--Gullstrand form, the metric becomes \(\mathrm{d}s^2 = - \mathfrak{f}(r)\mathrm{d}t^2 + 2\mathfrak{h}(r) \mathrm{d}t \mathrm{d}r + \mathrm{d}r^2 + r^2\mathrm{d}\Omega^2\), where \(\mathfrak{h}(r) = \sqrt{\mathfrak{f}(r)\big(\mathfrak{g}(r)^{-1} - 1\big)}\). The tunneling rate is linked to the imaginary part of the action \cite{parikh2004energy, vanzo2011tunnelling, calmet2023quantum}. 

The action \(\mathcal{S}\) for a particle moving freely in curved spacetime can be written as \(\mathcal{S} = \int p_\mu \, \mathrm{d}x^\mu\). When computing \(\text{Im} \, \mathcal{S}\), only the first term in \(p_\mu \mathrm{d}x^\mu = p_t \mathrm{d}t + p_r \mathrm{d}r\) contributes, as \(p_t \mathrm{d}t = -\omega \mathrm{d}t\) remains real, leaving the imaginary component unaffected. Therefore,
\ie
\text{Im}\,\mathcal{S}=\text{Im}\,\int_{r_i}^{r_f} \,p_r\,\mathrm{d}r=\text{Im}\,\int_{r_i}^{r_f}\int_{0}^{p_r} \,\mathrm{d}p_r'\,\mathrm{d}r.
\fe
Using Hamilton's equations for a system with Hamiltonian \(H = M - \omega'\), we find that \(\mathrm{d}H = -\mathrm{d}\omega'\), where \(0 \leq \omega' \leq \omega\) and \(\omega\) denotes the energy of the emitted particle. Consequently, this yields:
\ie
\begin{split}
\text{Im}\, \mathcal{S} & = \text{Im}\,\int_{r_i}^{r_f}\int_{M}^{M-\omega} \,\frac{\mathrm{d}H}{\mathrm{d}r/\mathrm{d}t}\,\mathrm{d} r  =\text{Im}\,\int_{r_i}^{r_f}\,\mathrm{d}r\int_{0}^{\omega} \,-\frac{\mathrm{d}\omega'}{\mathrm{d}r/\mathrm{d}t}\,.
\end{split}
\fe
By rearranging the order of integration and performing the substitution
\ie
    \frac{\mathrm{d}r}{\mathrm{d}t} = -\mathfrak{h}(r)+\sqrt{\mathfrak{f}(r)+\mathfrak{h}(r)^2}=1-\sqrt{\frac{\Delta(r)}{r}}, 
    \fe
where $\Delta(r)= 2(M- \omega^{\prime}) - \frac{\ell}{1-\ell}r$. In this manner, it reads
\ie
\label{ims}
\text{Im}\, \mathcal{S} =\text{Im}\,\int_{0}^{\omega} -\mathrm{d}\omega'\int_{r_i}^{r_f}\,\frac{\mathrm{d}r}{1-\sqrt{\frac{\Delta(r,\,\omega^\prime)}{r}}}.
\fe

\begin{figure}
    \centering
      \includegraphics[scale=0.55]{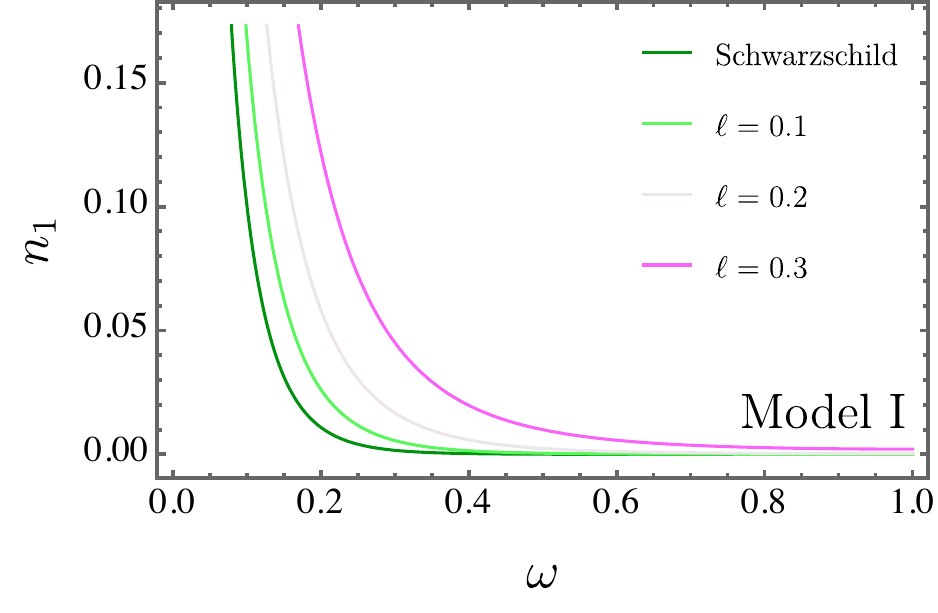}
    \caption{The particle density \( n_{1} \) is shown for different values of \( \ell \) in Model I.}
    \label{particledensitybosons1}
\end{figure}

With \(M\) replaced by \((M - \omega')\) in the original metric, the function \(\Delta(r)\) now depends on \(\omega'\). This adjustment introduces a pole at the new horizon, \(r = \mathfrak{r}\). Performing a contour integration around this point in the counterclockwise direction results in:
\begin{eqnarray}
    \text{Im}\, \mathcal{S}  = 4\pi (1-\ell)^{2} \omega \left( M - \frac{\omega}{2} \right)  .
\end{eqnarray}
According to \cite{vanzo2011tunnelling}, the emission rate for a \textit{Hawking} particle, incorporating Lorentz--violating corrections, can be written as
\ie
\Gamma \sim e^{-2 \, \text{Im}\, S}=e^{-8 \pi (1-\ell)^{2} \omega \left( M - \frac{\omega}{2} \right)} .
\fe
Notably, in the limit where \(\omega \to 0\), the emission spectrum reverts to the standard Planckian form originally obtained by Hawking. Accordingly, the emission spectrum is represented by
\begin{equation}
    \mathfrak{P}(\omega)=\frac{\mathrm{d}\omega}{2\pi}\frac{1}{e^{8 \pi (1-\ell)^{2} \omega \left( M - \frac{\omega}{2} \right)
    }-1}.
\end{equation}
The emission spectrum, with its added dependence on \(\omega\), departs from the conventional blackbody distribution, a result that becomes evident upon inspection. For low \(\omega\) values, the spectrum reduces to a Planck--like form but with an adjusted \textit{Hawking} temperature. Furthermore, the particle number density can be formulated in terms of the tunneling rate
\ie
n_{1} = \frac{\Gamma}{1 - \Gamma} = \frac{1}{e^{8 \pi  (1-\ell)^2 \omega  \left(M-\frac{\omega }{2}\right)} - 1}.
\fe
To clarify the behavior of \(n\), Fig. \ref{particledensitybosons1} illustrates its dependence on the Lorentz--violating parameter \(\ell\). The figure confirms that an increase in \(\ell\) leads to a higher particle number density. Additionally, \(n_{1}\) is compared to the Schwarzschild case for reference.

In essence, these results suggest that radiation emitted from a black hole encodes information about its internal state. The \textit{Hawking} amplitudes are affected by the Lorentz--violating parameter \(\ell\), and the power spectrum incorporates these corrections, diverging from the traditional blackbody spectrum when energy conservation is taken into account.


\subsection{Fermionic modes}

Black holes, due to their intrinsic temperature, emit radiation akin to black body radiation, though this emission excludes adjustments for \textit{greybody} factors. The resulting spectrum is predicted to contain particles of various spin types, including fermions. Studies by Kerner and Mann \cite{o69}, along with further investigations \cite{o75,o72,o71,o74,o73,o70}, indicate that both massless bosons and fermions emerge with the same temperature. Furthermore, research on spin--$1$ bosons has demonstrated that even when higher--order quantum corrections are applied, the \textit{Hawking} temperature remains stable \cite{o77,o76}.

For fermions, the action is commonly represented by the phase of the spinor wave function, which satisfies the Hamilton--Jacobi equation. An alternative expression for the action is given by \cite{o83, o84, vanzo2011tunnelling}
\[ S_{f} = S_{0} + (\text{spin corrections}), \]
where \( S_0 \) denotes the classical action for scalar particles. The spin corrections account for the coupling between the particle’s spin and the spin connection of the manifold, yet they do not induce singularities at the horizon. It should be noted that these effects are minimal, impacting primarily the precession of spin, and can thus be disregarded here. Additionally, any impact on the black hole’s angular momentum from emitted particle spins is negligible, especially for black holes without rotation and with masses considerably larger than the Planck scale \cite{vanzo2011tunnelling}. Statistically, the symmetric emission of particles with opposing spins ensures that the black hole's angular momentum remains unchanged.

\begin{figure}
    \centering
      \includegraphics[scale=0.55]{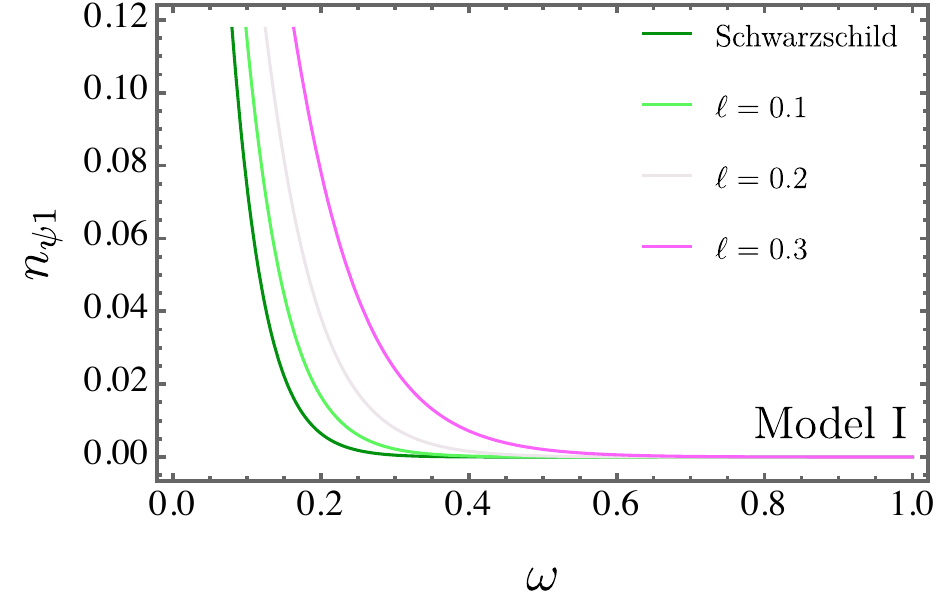}
    \caption{The particle density \( n_{\psi 1} \) is shown for various values of \( \ell \).}
    \label{particledensityfermions}
\end{figure}

Expanding upon our previous findings, we investigate the tunneling of fermionic particles as they cross the event horizon of the specific black hole model. The emission rate is calculated within a Schwarzschild--like coordinate system, known for its singularity at the horizon. For complementary analyses utilizing generalized Painlevé--Gullstrand and Kruskal--Szekeres coordinates, refer to the foundational study \cite{o69}. As a starting point, we introduce a general metric expressed by:
$\mathrm{d}s^{2} = \mathfrak{A}(r) \mathrm{d}t^{2} + [1/\mathfrak{B}(r)]\mathrm{d}r^{2} + \mathfrak{C}(r)[\mathrm{d}\theta^{2} + r^{2}\sin^{2}\theta ]\mathrm{d}\varphi^{2}$.
In curved spacetime, the Dirac equation is formulated as
$
\left(\gamma^\mu \nabla_\mu + \frac{m}{\hbar}\right) \Psi(t,r,\theta,\varphi) = 0$
with
$
\nabla_\mu = \partial_\mu + \frac{i}{2} {\Gamma^\alpha_{\;\mu}}^{\;\beta} \,\Sigma_{\alpha\beta}$ and $ 
\Sigma_{\alpha\beta} = \frac{i}{4} [\gamma_\alpha,  \gamma_\beta]$.
The \( \gamma^\mu \) matrices fulfill the conditions of the Clifford algebra, expressed by the relation
$
\{\gamma_\alpha,\gamma_\beta\} = 2 g_{\alpha\beta} \mathbb{1}$, 
where \(\mathbb{1}\) denotes the \(4 \times 4\) identity matrix. In this context, we choose the \(\gamma\) matrices in the following form:
\begin{eqnarray*}
 \gamma ^{t} &=&\frac{i}{\sqrt{\mathfrak{A}(r)}}\left( \begin{array}{cc}
\vec{1}& \vec{ 0} \\ 
\vec{ 0} & -\vec{ 1}%
\end{array}%
\right), \;\;
\gamma ^{r} =\sqrt{\mathfrak{B}(r)}\left( 
\begin{array}{cc}
\vec{0} &  \vec{\sigma}_{3} \\ 
 \vec{\sigma}_{3} & \vec{0}%
\end{array}%
\right), \\
\gamma ^{\theta } &=&\frac{1}{r}\left( 
\begin{array}{cc}
\vec{0} &  \vec{\sigma}_{1} \\ 
 \vec{\sigma}_{1} & \vec{0}%
\end{array}%
\right), \;\;
\gamma ^{\varphi } =\frac{1}{r\sin \theta }\left( 
\begin{array}{cc}
\vec{0} &  \vec{\sigma}_{2} \\ 
 \vec{\sigma}_{2} & \vec{0}%
\end{array}%
\right),
\end{eqnarray*}%
where \(\vec{\sigma}\) denotes the Pauli matrices, complying with the conventional commutation:
$
 \sigma_i  \sigma_j = \vec{1} \delta_{ij} + i \varepsilon_{ijk} \sigma_k, \,\, \text{in which}\,\, i,j,k =1,2,3. 
$ The matrix for $\gamma^{5}$ is instead
\begin{equation*}
\gamma ^{5}=i\gamma ^{t}\gamma ^{r}\gamma ^{\theta }\gamma ^{\varphi }=i\sqrt{\frac{\mathfrak{B}(r)}{\mathfrak{A}(r)}}\frac{1}{r^{2}\sin \theta }\left( 
\begin{array}{cc}
\vec{ 0} & - \vec{ 1} \\ 
\vec{ 1} & \vec{ 0}%
\end{array}%
\right)\:.
\end{equation*}
To describe a Dirac field with spin oriented upward along the positive \(r\)--axis, we adopt the following ansatz \cite{vagnozzi2022horizon}:
\begin{equation}
\Psi_{\uparrow}(t,r,\theta ,\varphi ) = \left( \begin{array}{c}
\mathfrak{H}(t,r,\theta ,\varphi ) \\ 
0 \\ 
\mathfrak{Y}(t,r,\theta ,\varphi ) \\ 
0%
\end{array}%
\right) \exp \left[ \frac{i}{\hbar }\psi_{\uparrow}(t,r,\theta ,\varphi )\right]\;.
\label{spinupbh} 
\end{equation} 
We focus specifically on the spin--up (\(\uparrow\)) case, while noting that the spin--down (\(\downarrow\)) case, oriented along the negative \(r\)--axis, follows a similar process. Substituting the ansatz (\ref{spinupbh}) into the Dirac equation results in:
\ie
\begin{split}
-\left( \frac{i \,\mathfrak{H}}{\sqrt{\mathfrak{A}(r)}}\,\partial _{t} \psi_{\uparrow} + \mathfrak{Y} \sqrt{\mathfrak{B}(r)} \,\partial_{r} \psi_{\uparrow}\right) + \mathfrak{H} m &=0, \\
-\frac{\mathfrak{Y}}{r}\left( \partial _{\theta }\psi_{\uparrow} +\frac{i}{\sin \theta } \, \partial _{\varphi }\psi_{\uparrow}\right) &= 0, \\
\left( \frac{i \,\mathfrak{Y}}{\sqrt{\mathfrak{A}(r)}}\,\partial _{t}\psi_{\uparrow} - \mathfrak{H} \sqrt{\mathfrak{B}(r)}\,\partial_{r}\psi_{\uparrow}\right) + \mathfrak{Y} m &= 0, \\
-\frac{\mathfrak{H}}{r}\left(\partial _{\theta }\psi_{\uparrow} + \frac{i}{\sin \theta }\,\partial _{\varphi }\psi_{\uparrow}\right) &= 0,
\end{split}
\fe%
to leading order in \(\hbar\). We consider the action to be expressed as
$
\psi_{\uparrow}=- \omega\, t + \chi(r) + L(\theta ,\varphi )  $
so that the equations below are brought about
\cite{vanzo2011tunnelling}
\begin{eqnarray}
\left( \frac{i\, \omega\, \mathfrak{H}}{\sqrt{\mathfrak{A}(r)}} - \mathfrak{Y} \sqrt{\mathfrak{B}(r)}\, \mathfrak \chi^{\prime }(r)\right) +m\, \mathfrak{H} &=&0,
\label{bhspin5} \\
-\frac{\mathfrak{H}}{r}\left( L_{\theta }+\frac{i}{\sin \theta }L_{\varphi }\right) &=&0,
\label{bhspin6} \\
-\left( \frac{i\,\omega\, \mathfrak{Y}}{\sqrt{\mathfrak{A}(r)}} + \mathfrak{H}\sqrt{\mathfrak{B}(r)}\, \mathfrak \chi^{\prime }(r)\right) +\mathfrak{Y}\,m &=&0,
\label{bhspin7} \\
-\frac{\mathfrak{H}}{r}\left( L_{\theta } + \frac{i}{\sin \theta }L_{\varphi }\right) &=& 0.
\label{bhspin8}
\end{eqnarray}
The forms of \(\mathfrak{H}\) and \(\mathfrak{Y}\) do not affect the outcome that Equations (\ref{bhspin6}) and (\ref{bhspin8}) require \(L_{\theta} + i(\sin \theta)^{-1} L_{\varphi} = 0\), meaning \(L(\theta, \varphi)\) must be complex. This condition for \(L\) holds true for both outgoing and incoming cases. Thus, when calculating the ratio of outgoing to incoming probabilities, the contributions from \(L\) cancel, allowing us to omit \(L\) in further steps. In the massless scenario, Eqs. (\ref{bhspin5}) and (\ref{bhspin7}) provide two possible solutions:
\ie
\mathfrak{H} = -i \mathfrak{Y}, \qquad \chi^{\prime }(r) = \chi_{\text{out}}' = \frac{\omega}{\sqrt{\mathfrak{A}(r)\mathfrak{B}(r)}},
\fe
\ie
\mathfrak{H} = i \mathfrak{Y}, \qquad \chi^{\prime }(r) = \chi_{\text{in}}' = - \frac{\omega}{\sqrt{\mathfrak{A}(r)\mathfrak{B}(r)}},
\fe
where \(\chi_{\text{out}}\) and \(\chi_{\text{in}}\) describe the outgoing and incoming solutions, respectively \cite{vanzo2011tunnelling}. Consequently, the overall tunneling probability is given by \(\Gamma_{\psi} \sim e^{-2 \, \text{Im} \, (\chi_{\text{out}} - \chi_{\text{in}})}\). Thereby,
\begin{equation}
\mathcal \chi_{ \text{out}}(r)= - \mathcal \chi_{ \text{in}} (r) = \int \mathrm{d} r \,\frac{\omega}{\sqrt{\mathfrak{A}(r)\mathfrak{B}(r)}}\:.
\end{equation}%
Notably, given the dominant energy condition and the Einstein field equations, the functions \(\mathfrak{A}(r)\) and \(\mathfrak{B}(r)\) share the same zeros. Thus, near \(r = \mathfrak{r}\), to first order, we can express this as:
\ie
\mathfrak{A}(r)\mathfrak{B}(r) = \mathfrak{A}'(\mathfrak{r})\mathfrak{B}'(\mathfrak{r})(r - \mathfrak{r})^2 + \dots
\fe
showing the presence of a simple pole with a specific coefficient. Using Feynman’s method, we obtain
\ie
2\mbox{ Im}\;\left( \mathcal \chi_{ \text{out}} - \mathcal \chi_{ \text{in}} \right) =\mbox{ Im}\int \mathrm{d} r \,\frac{4\omega}{\sqrt{\mathfrak{A}(r)\mathfrak{B}(r)}}=\frac{2\pi\omega}{\kappa},
\fe
with the surface gravity defined as \(\kappa = \frac{1}{2} \sqrt{\mathfrak{A}'(r_{h}) \mathfrak{B}'(r_{h})}\). Under these conditions, the expression \(\Gamma_{\psi} \sim e^{-\frac{2 \pi \omega}{\kappa}}\) determines the particle density \(n_{\psi}\) for this black hole solution
\ie
n_{\psi1} = \frac{\Gamma_{\psi}}{1+\Gamma_{\psi}}  = \frac{1}{e^{8 \pi  (1- \ell)^2 M \omega }+1}.
\fe
In Fig. \ref{particledensityfermions}, we illustrate the variation of \(n_{\psi1}\) across different values of \(\ell\), alongside a comparison with the standard Schwarzschild case. In general lines, an increase in \(\ell\) corresponds to a rise in particle density. Compared to the Lorentz--violating influenced curves, the Schwarzschild case appears as the lowest curve.

\subsection{Greybody factors for bosons}


For the sake of performed a general analysis, let us consider a metric of the following form as we did the previous sections:
\ie
\nonumber
\mathrm{d}s^{2} = \mathcal{A}(r) \mathrm{d}t^{2} + [1/\mathcal{B}(r)]\mathrm{d}r^{2} + \mathcal{C}(r)[\mathrm{d}\theta^{2} + r^{2}\sin^{2}\theta ]\mathrm{d}\varphi^{2}.
\fe

Starting from the Klein--Gordon equation in a spherically symmetric curved spacetime, the partial wave equation emerges. Through the application of the separation of variables technique, the equation is transformed into a more simplified representation
\ie
\label{sai}
{\Psi _{\omega lm}}(r,t) = \frac{{{\psi_{\omega l}}(r)}}{r}{Y_{lm}}(\theta ,\varphi ){e^{ - i \omega t}}.
\fe
For a spherically symmetric spacetime, the tortoise coordinate $r^{*}$ is defined in terms of the metric components corresponding to the temporal and radial coordinates, expressed as
\ie
{\rm{d}}r^{*} = \frac{{\rm{d}}r}{\sqrt {\mathcal{A}(r)\mathcal{B}(r)} },
\fe
in which this redefinition modifies the Klein--Gordon equation into a form analogous to the Schrödinger wave equation.
\ie
\left[\frac{{{\rm{d}^2}}}{{\mathrm{d}{{r^*}^2}}} + ({\omega ^2} - \mathcal{V}_{1})\right]\psi_{\omega l} (r) = 0.
\fe
The term $\mathcal{V}_{1}$ describes the effective potential influencing scalar perturbations, formulated as
\ie
\begin{split}
\mathcal{V}_{1} & = \mathcal{A}(r)\left[\frac{{l(l + 1)}}{{{r^2}}} + \frac{1}{{r\sqrt {{\mathcal{A}(r)}{\mathcal{B}(r)^{ - 1}}} }}\frac{\mathrm{d}}{{\mathrm{d}r}}\sqrt {{\mathcal{A}(r)}\mathcal{B}(r)}\right]\\
& = \frac{-\frac{l (l+1) r}{\ell-1}-2 l (l+1) M+2 M \sqrt{\frac{(2 (\ell-1) M+r)^2}{(\ell - 1)^2 r^2}}}{r^3}.
\end{split}
\fe

\begin{figure}
    \centering
      \includegraphics[scale=0.55]{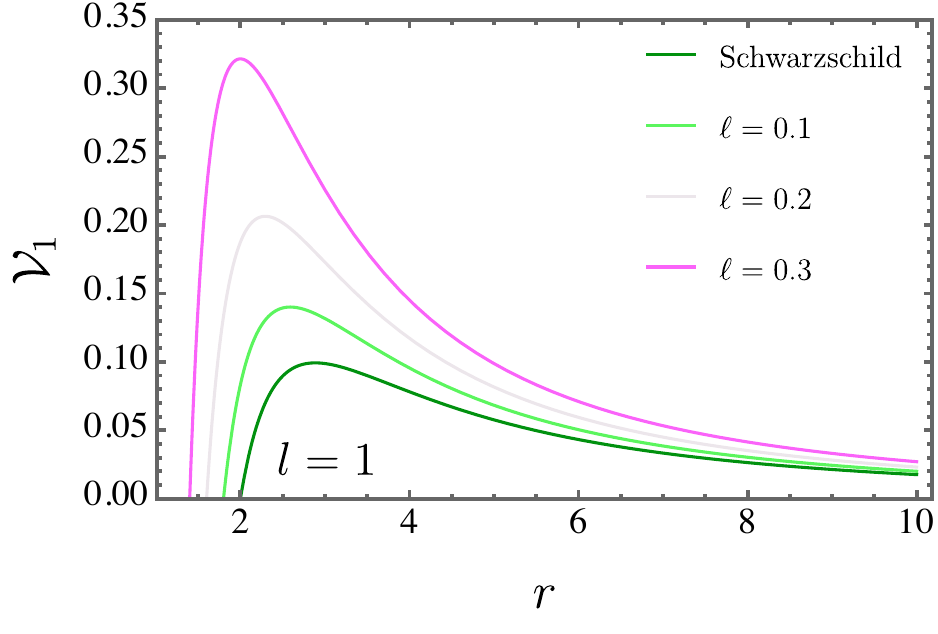}
    \caption{The effective potential $\mathcal{V}_{1}$ is presented for various values of $\ell$ while keeping $l = 1$ fixed. For reference, the Schwarzschild case is also included in the comparison.}
    \label{potentialmodelI}
\end{figure}

To corroborate our outcomes, we present Fig. \ref{potentialmodelI}. In general lines, it shows the effective potential $\mathcal{V}_{1}$ for scalar perturbations. As the parameter $\ell$ increases, the potential heightens accordingly. The Schwarzschild case, characterized by the lowest potential, is included for reference.

In addition, the propagation of Hawking radiation from the event horizon to infinity is influenced by spacetime curvature, causing deviations from a perfect blackbody spectrum. This deviation is described by the greybody factor. Here, it for a massless spin--$0$ field is analyzed using semi--analytic approaches \cite{sakalli2022topical,boonserm2019greybody,ovgun2024shadow,al2024fermionic,boonserm2008transmission}. The upper bound of the greybody factor, denoted by $T_{b1}$, is expressed as
\ie
\label{tgmetric}
T_{b1} \ge {\mathop{\rm sech}\nolimits} ^2 \left(\int_\infty^ {+\infty} {\mathfrak{G} \,\rm{d}}r^{*}\right),
\fe
where
\ie
\mathfrak{G} = \frac{{\sqrt {{{(\xi')}^2} + {{({\omega ^2} - \mathcal{V}_{1} - {\xi^2})}^2}} }}{{2\xi}}.
\fe
The positive function $\xi$ is constrained by the boundary conditions $\xi(+\infty) = \xi(-\infty) = \omega$. Fixing $\xi$ to the constant value $\omega$ leads to a simplified form of Eq. \eqref{tgmetric}
\ie
\begin{split}
& T_{b1}  \ge {\mathop{\rm sech}\nolimits} ^2 \left[\int_{-\infty}^ {+\infty} \frac{\mathcal{V}_{1}} {2\omega}\mathrm{d}r^{*}\right] \\
& ={\mathop{\rm sech}\nolimits} ^2 \left[\int_{r_{ h}}^ {+\infty} \frac{\mathcal{V}_{1}} {2\omega\sqrt{\mathcal{A}(r)\mathcal{B}(r)}}\mathrm{d} r\right] \\
& ={\mathop{\rm sech}\nolimits} ^2 \left[ \frac{1}{2\omega} \left(\frac{1 + 2 l (l+1) (1 - \ell)}{4 (1 - \ell )^2 M} \right)  \right].
\end{split}
\fe

\begin{figure}
    \centering
      \includegraphics[scale=0.51]{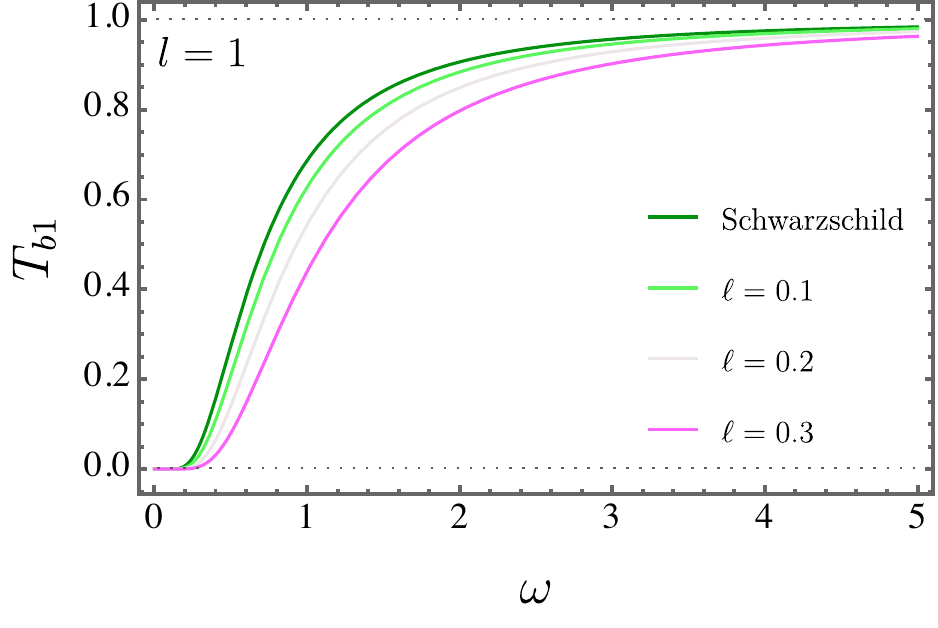}
      \includegraphics[scale=0.51]{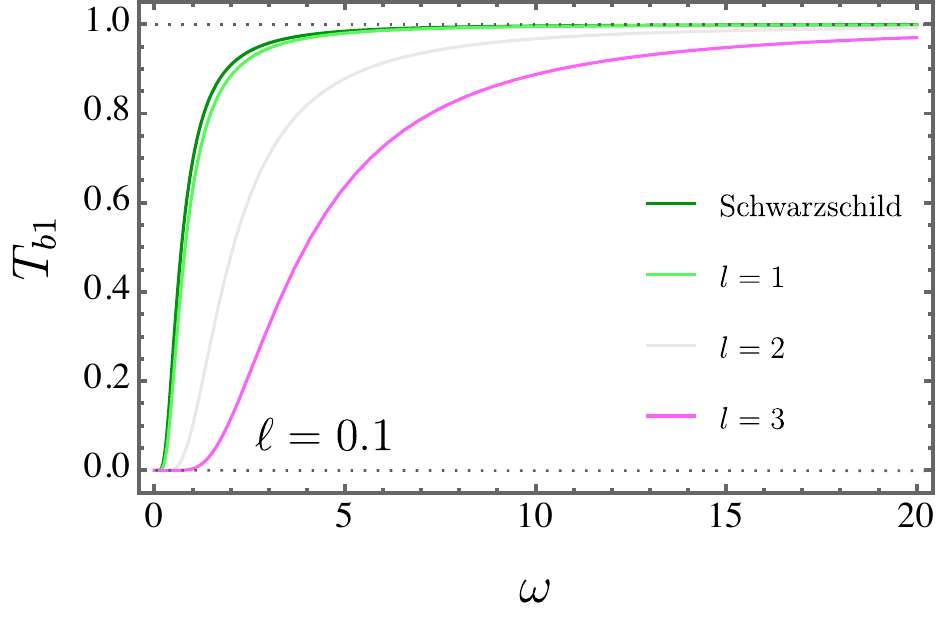}
    \caption{The greybody factors $T_{b1}$ are shown for varying $\ell$ with $l=1$ fixed (left panel) and for varying $l$ with $\ell$ held constant at 0.1 (right panel). In both scenarios, the results are compared with the Schwarzschild case.}
    \label{greybodybosonsmodelI}
\end{figure}

Fig. \ref{greybodybosonsmodelI} displays the greybody factors for bosons, $T_{b1}$, under two conditions: the left panel shows variations in $\ell$ with $l=1$ fixed, while the right panel explores changes in $l$ while keeping $\ell = 0.1$ constant. Both cases are compared against the Schwarzschild solution for reference. In general, for these configurations, increasing $\ell$ and $l$ leads to a reduction in the greybody factors $T_{b1}$.

\subsection{Greybody factors for fermions}

This analysis investigates the dynamics of massless Dirac perturbations in a static, spherically symmetric black hole spacetime. The Newman--Penrose formalism is employed as the framework to describe the evolution of the massless spin--1/2 field. The Dirac equations governing the system are expressed as \cite{newman1962approach, chandrasekhar1984mathematical}:
\begin{align}
(D + \epsilon - \rho) \psi_1 +( \bar{\delta} + \pi - \alpha) \psi_2 &= 0, \\
(\Delta + \mu - \gamma) \psi_2 + (\delta + \beta - \tau) \psi_1 &= 0.
\end{align}

The Dirac spinors $\psi_1$ and $\psi_2$ are defined in terms of directional derivatives constructed from a chosen null tetrad. These derivatives include $D = l^\mu \partial_\mu$, $\Delta = n^\mu \partial_\mu$, $\delta = m^\mu \partial_\mu$, and $\bar{\delta} = \bar{m}^\mu \partial_\mu$, where each component is associated with a specific tetrad vector.

Next, the null tetrad basis vectors are determined from the metric components and can be expressed in the form:
\ie
\begin{split}
l^\mu &= \left(\frac{1}{\mathcal{A}(r)}, \sqrt{\frac{\mathcal{B}(r)}{\mathcal{A}(r)}}, 0, 0\right), \\
n^\mu & = \frac{1}{2} \left(1, -\sqrt{\mathcal{A}(r) \mathcal{B}(r)}, 0, 0\right), \\
m^\mu &= \frac{1}{\sqrt{2} r} \left(0, 0, 1, \frac{i}{\sin \theta}\right), \\
\bar{m}^\mu &= \frac{1}{\sqrt{2} r} \left(0, 0, 1, \frac{-i}{\sin \theta}\right).
\end{split}
\fe

From these definitions, the spin coefficients with nonzero values can be expressed as
\ie
\begin{split}
 & \rho = -\frac{1}{r} \frac{\mathcal{B}(r)}{\mathcal{A}(r)},  \quad
\mu = -\frac{\sqrt{\mathcal{A}(r) \mathcal{B}(r)}}{2r}, \\ & \gamma = \frac{\mathcal{A}(r)'}{4}\sqrt{\frac{\mathcal{B}(r)}{\mathcal{A}(r)}},  \quad
\beta = -\alpha = \frac{\cot{\theta}}{2\sqrt{2}r}. 
\end{split}
\fe

Isolating the equations governing the dynamics of a massless Dirac field leads to a single equation for $\psi_1$, effectively describing its behavior
\begin{align}
\left[(D - 2\rho)(\Delta + \mu - \gamma) - (\delta + \beta) (\bar{\delta}+\beta)\right] \psi_1 = 0.
\end{align}

Incorporating the explicit expressions for the directional derivatives and spin coefficients allows the equation to be reformulated below
\ie
\begin{split}
&\left[ \frac{1}{2\mathcal{A}(r)} \partial_t^2 - \left( \frac{\sqrt{\mathcal{A}(r)\mathcal{B}(r)}}{2r} +\frac{\mathcal{A}(r)'}{4}\sqrt{\frac{\mathcal{B}(r)}{\mathcal{A}(r)}}\right)\frac{1}{\mathcal{A}(r)}\partial_t \right. \\
& \left. - \frac{\sqrt{\mathcal{A}(r)\mathcal{B}(r)}}{2} \sqrt{\frac{\mathcal{B}(r)}{\mathcal{A}(r)}}\partial_r^2 \right. \\
& \left. -\sqrt{\frac{\mathcal{B}(r)}{\mathcal{A}(r)}} \partial_r \left( \frac{\sqrt{\mathcal{A}(r)\mathcal{B}(r)}}{2} + \frac{\mathcal{A}(r)'}{4}{\sqrt{\frac{\mathcal{B}(r)}{\mathcal{A}(r)}}} \right) \right] \psi_1 \\ + 
&\left[ \frac{1}{\sin^2\theta} \partial_\phi^2 + i \frac{\cot \theta}{\sin \theta}\partial_\phi \right. \\
& \left.+ \frac{1}{\sin \theta}\partial_\theta \left( \sin \theta \partial_\theta \right) - \frac{1}{4} \cot^2 \theta + \frac{1}{2} \right] \psi_1 = 0.
\end{split}
\fe

To separate the equations into radial and angular components, the wave function turns out
\begin{align}
\psi_1 = \Psi(r) Y_{lm}(\theta, \phi) e^{-i \omega t},
\end{align}
so that
\begin{align}
&\left[  \frac{-\omega^2}{2\mathcal{A}(r)} - \left(\frac{\sqrt{\mathcal{A}(r)\mathcal{B}(r)}}{2r}+\frac{\mathcal{A}(r)'}{4} + \sqrt{\frac{\mathcal{B}(r)}{\mathcal{A}(r)}}\right)\frac{- i\omega}{\mathcal{A}(r)} \right. \\
& \left. - \frac{\sqrt{\mathcal{A}(r)\mathcal{B}(r)}}{2} \sqrt{\frac{\mathcal{B}(r)}{\mathcal{A}(r)}}\partial_r^2 -\lambda_{lm} \right. \\
& \left. - \sqrt{\frac{\mathcal{B}(r)}{\mathcal{A}(r)}}\partial_r \left(\frac{\sqrt{\mathcal{A}(r)\mathcal{B}(r)}}{2r} + \frac{\mathcal{A}(r)'}{4}\sqrt{\frac{\mathcal{B}(r)}{\mathcal{A}(r)}}\right) \right] \Psi(r) = 0.
\end{align}

The parameter $\lambda_{lm}$ acts as a separation constant in this formulation. Applying the generalized tortoise coordinate $r^*$ allows the radial wave equation to be reformulated into a Schrödinger--like structure, described by:
\begin{align}
\left[\frac{\mathrm{d}^2 }{\mathrm{d}r_*^2} +( \omega^2 - \mathcal{V}_{1 \psi}^{\pm}) \right]\Psi_{\pm}(r) = 0.
\end{align}
The potentials $\mathcal{V}_{1 \psi}^{\pm}$ corresponding to the massless spin--1/2 field are described in the form \cite{albuquerque2023massless, al2024massless, arbey2021hawking}:
\ie
\begin{split}\label{Vpm}
&\mathcal{V}_{1 \psi}^{\pm} = \frac{(l + \frac{1}{2})^2}{r^2} \mathcal{A}(r)  \pm \left(l + \frac{1}{2}\right) \sqrt{\mathcal{A}(r) \mathcal{B}(r)} \partial_r \left(\frac{\sqrt{\mathcal{A}(r)}}{r}\right) \\
& = \frac{\left(l+\frac{1}{2}\right)^2 \left(\frac{1}{1 - \ell} - \frac{2 M}{r}\right)}{r^2}   +   \frac{(2 l+1) (3 (1 - \ell) M - r) \sqrt{\frac{(- 2 (1 - \ell ) M+r)^2}{(1 - \ell)^2 r^2}}}{2 (1 - \ell) r^3 \sqrt{\frac{1}{1-\ell}-\frac{2 M}{r}}}.
\end{split}
\fe

\begin{figure}
    \centering
      \includegraphics[scale=0.55]{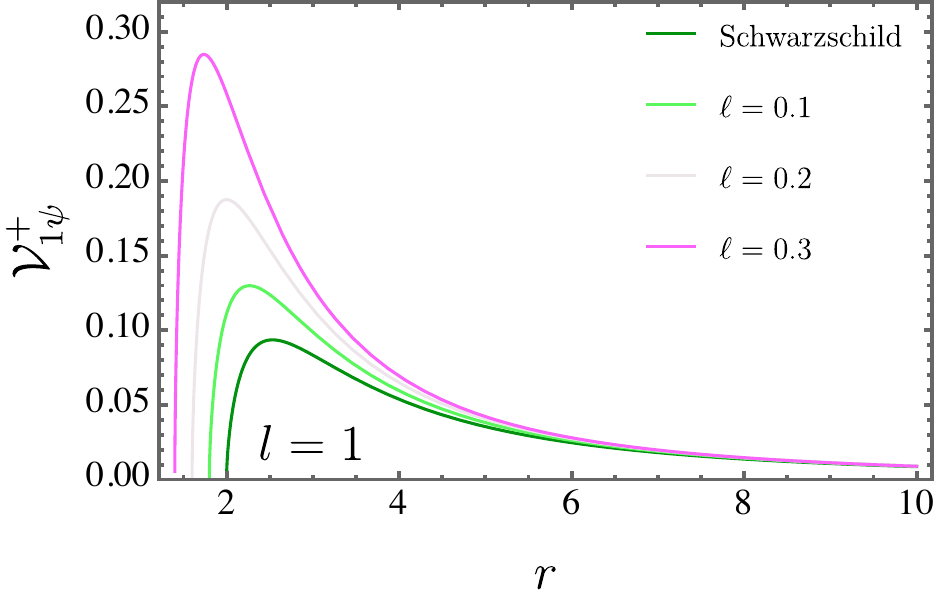}
    \caption{The effective potential for fermions, $\mathcal{V}_{1 \psi}^{+}$, is depicted for different choices of $\ell$ while keeping $l$ constant. The Schwarzschild case is also examined for reference.}
    \label{potentialmodelIfermions}
\end{figure}

In this analysis, the potential $\mathcal{V}_{1 \psi}^{+}$ is selected without any loss of generality, as a similar procedure applies to $\mathcal{V}_{1 \psi}^{-}$. Since the qualitative behavior of $\mathcal{V}_{1 \psi}^{-}$ closely mirrors that of $\mathcal{V}_{1 \psi}^{+}$ \cite{albuquerque2023massless,devi2020quasinormal}, the focus will remain on $\mathcal{V}_{1 \psi}^{+}$. Fig. \ref{potentialmodelIfermions} illustrates the features of $V^{+}_{1\psi}$, which, as expected, tends to zero as $r \to \infty$.

Moreover, applying the Dirac effective potential from Eq. \eqref{Vpm}, the greybody factor bounds for the bumblebee model within the metric formalism are obtained. These bounds can be written in a simplified form as:
\ie
\begin{split}
	T_{b1\psi} & \ge \mathrm{sech}^2 {\left(\frac{1}{2\omega}\int_{2M}^ {+\infty} \frac{\mathcal{V}_{1 \psi}^{+}} {\sqrt{\mathcal{A}(r)\mathcal{B}(r)}} \rm{d}r\right) }\\
 & =  \mathrm{sech}^2 \left[\frac{1}{2\omega} \left(  \frac{4 l^2+4 l+1}{2 (4 M-4 \ell M)} \right)\right].
\end{split}
\fe

\begin{figure}
    \centering
      \includegraphics[scale=0.51]{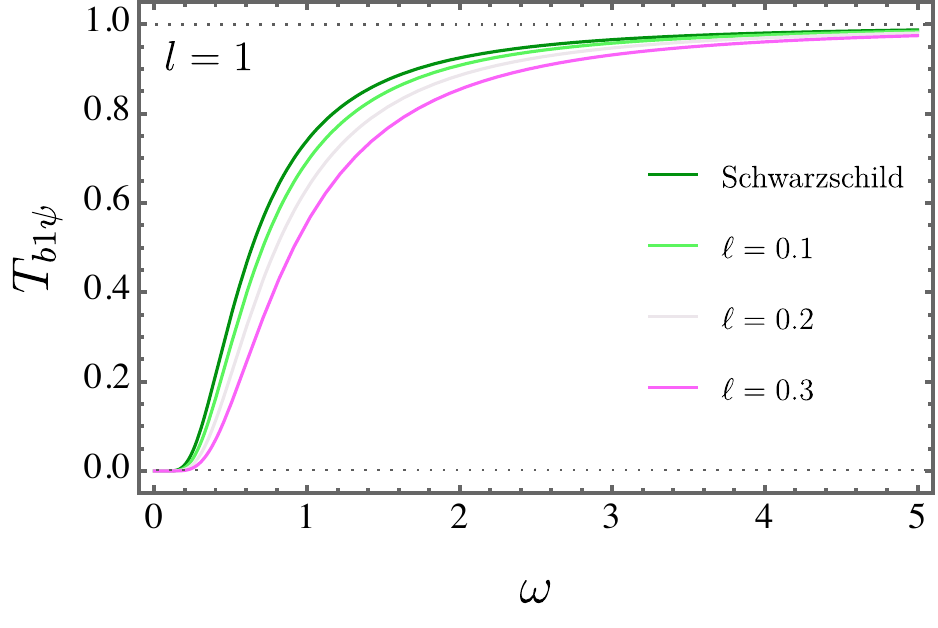}
      \includegraphics[scale=0.51]{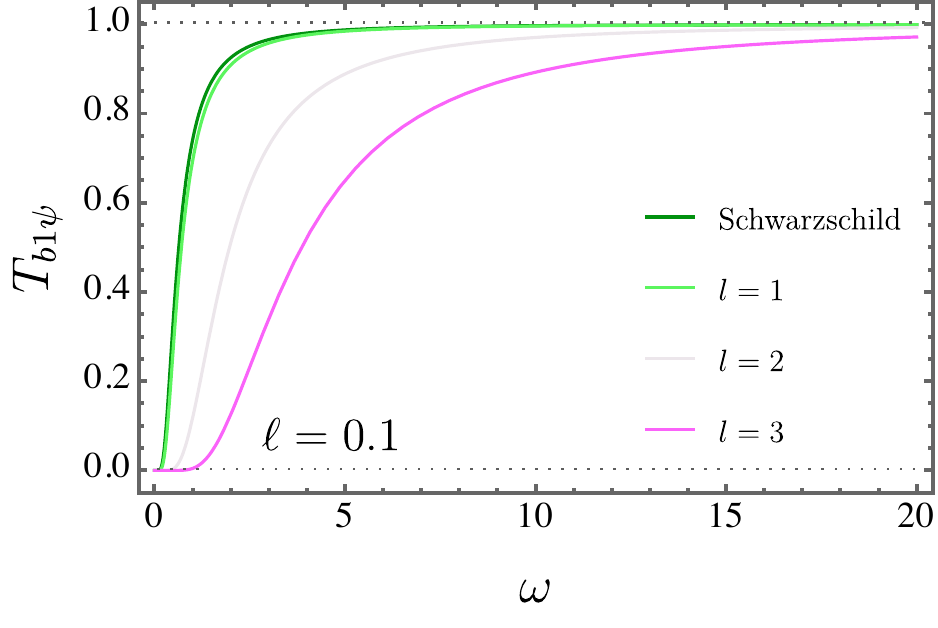}
    \caption{The greybody factors for fermions, $T_{b1\psi}$, are shown for varying values of $\ell$ with $l=1$ fixed (top panel) and for varying $l$ with $\ell$ set to 0.1. In both scenarios, the Schwarzschild case is included for comparison.}
    \label{greybodyfermionsmodelI}
\end{figure}

Fig. \ref{greybodyfermionsmodelI} presents the behavior of the greybody factor as a function of the frequency $\omega$. The left panel explores how $T_{b1\psi}$ responds to varying values of the Lorentz symmetry--breaking parameter $\ell$ while keeping $M = 1$ and $l = 1$ fixed. The plots indicate that increasing $\ell$ results in a reduction of $T_{b1\psi}$. Furthermore, the greybody factor for Model I consistently remains lower than the Schwarzschild case. The right panel displays the variation of $T_{b1\psi}$ with respect to $\omega$ for different values of $l$ while maintaining $\ell = 0.1$ constant. As $l$ grows, $T_{b1\psi}$ also diminishes, with the Schwarzschild scenario consistently showing the highest magnitude.

\subsection{The evaporation process}

Here, we examine qualitatively the black hole evaporation process described by Eq. (\ref{model1}). To do so, we fundamentally employ the \textit{Stefan--Boltzmann} law \cite{ong2018effective}
\ie
\label{sflaw}
\frac{\mathrm{d}M}{\mathrm{d}t}  =  - \alpha a \sigma T^{4},
\fe
where \(\sigma\) denotes the cross--sectional area, \(a\) is the radiation constant, \(\alpha\) represents the \textit{greybody} factor and \(T\) represents the \textit{Hawking} temperature. It is fundamental to mention that the radiation primarily consists of massless photons and neutrinos \cite{hiscock1990evolution,page1976particle}. In other words, $\sigma$ is determined as $\pi \mathcal{R}^2$, with $\mathcal{R}$ representing the shadow radius. In addition, within this framework, the greybody factors reduce to $\alpha \to 1$ \cite{liang2025einstein}.

Within the geometric optics approximation, \(\sigma\) is identified as the photon capture cross section, defined as follows
\ie
\sigma = 27(1-\ell )^{3}M^{2}\pi.
\fe

The metric in question features a timelike Killing vector, represented as $\Tilde{\xi} = \partial / \partial t$, which allows for the definition of a conserved quantity. This conserved quantity arises from the symmetry encoded in $\Tilde{\xi}$ and can be derived through the application of the Killing equation, as outlined below:
\ie
\label{surface1}
\nabla^\nu(\Tilde{\xi}^\mu \Tilde{\xi}_\mu) = -2\kappa\Tilde{\xi}^\nu.
\fe
In this framework, $\nabla_\nu$ denotes the covariant derivative, while $\kappa$ is the surface gravity, which constant along the trajectories determined by $\Tilde{\xi}$. This constancy is expressed by the vanishing of the Lie derivative of $\kappa$ with respect to $\Tilde{\xi}$, as follows:
\ie
\label{surface2}
\mathcal{L}_{\Tilde{\xi}}\kappa = 0.
\fe
Significantly, $\kappa$ remains constant on the horizon. In the coordinate basis, the timelike Killing vector has components $\Tilde{\xi}^{\mu} = (1, 0, 0, 0)$. Thereby, we have
\ie 
\label{surface3}
\kappa = {\left.\frac{f^{\prime}(r)}{2} \right|_{r = {\mathfrak{r}^{(\mathrm{I})}}}},
\fe 
where $\mathfrak{r}^{(\mathrm{I})}$ represents the event horizon for the Model I, which is written as $\mathfrak{r}^{(\mathrm{I})} = 2M (1 - \ell)$.

\begin{figure}
    \centering
      \includegraphics[scale=0.55]{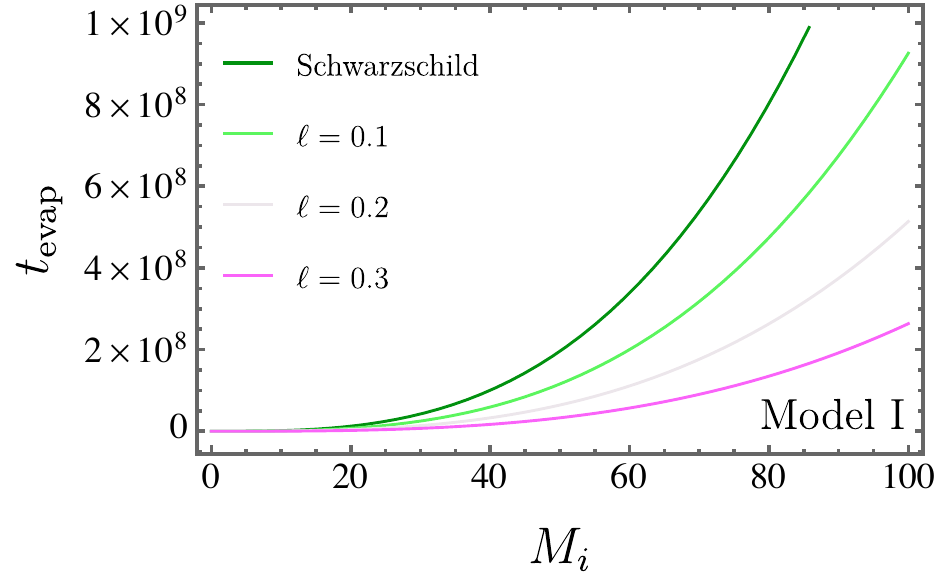}
    \caption{The evaporation time \( t_{\text{evap}} \) is shown for different values of \( \ell \).}
    \label{evap}
\end{figure}

The seminal work of Hawking, as outlined in Ref. \cite{hawking1975particle}, established that black holes emit radiation, which is characterized by the Hawking temperature. This temperature is defined as $ T = \kappa / 2\pi$. In the context of our analysis, it reads 
\ie
\label{haah}
T = \frac{1}{8 \pi  (1- \ell)^2 M}.
\fe
It is worth noting that the calculation of the Hawking temperature using the surface gravity aligns with the earlier result obtained through the analysis of Hawking radiation via the Bogoliubov coefficients, as expected. With all these preliminaries, Eq. (\ref{sflaw}) turns out to be 
\ie
\frac{\mathrm{d}M}{\mathrm{d}t} = \frac{27 \xi}{4096 \pi ^3 (\ell-1)^5 M^2},
\fe
with \(\xi = a \alpha\). The task now is to compute the integral below
\ie
\begin{split}
\int_{0}^{t_{\text{evap}}} \xi \mathrm{d}\tau & = - \int_{M_{i}}^{M_{f}} 
\left[ \frac{27 }{4096 \pi ^3 (1-\ell)^5 M^2}   \right]^{-1} \mathrm{d}M,
\end{split}
\fe
where \(t_{\text{evap}}\) represents the evaporation time. Accordingly, it is expressed as
\ie
t_{\text{evap}} = - \frac{4096 \pi ^3 (1-\ell)^5 \left(M_{f}^3 - M_{i}^3\right)}{81 \xi}.
\fe
Observe that \(t_{\text{evap}}\) is determined through analytical calculation. For the black hole in question, no remnant mass is expected; thus, we assume it will evaporate entirely, meaning \(M_{f} \to 0\), namely,
\ie
t_{\text{evap}} = \frac{4096 \pi ^3 (1-\ell)^5 M_{i}^3}{81 \xi}.
\fe
To provide a clearer interpretation of our findings, we plot \(t_{\text{evap}}\) in Fig. \ref{evap} for various values of \(\ell\). It is evident that as \(\ell\) increases, the evaporation time decreases. Notice that, consistent with the calculation presented in the previous subsection for the greybody factors, their magnitude tends to $1$. 

\bigskip 

\section{Model II}


\subsection{Bosonic modes}


\subsubsection{The \textit{Hawking} radiation}

Similarly to the approach taken in the previous section for Model I, here, for Model II (in particular, analogous to Eq. (\ref{rfgr}) in Model I), we express
\ie
\label{sddsaas}
r_{2} = 2 M-\frac{E \lambda }{\sqrt{1-\ell}},
\fe
where we also account for the negative solution of the square root in Eq. (\ref{eq:drdl}) to address ingoing geodesics. Thus, we express the solution of Eq. (\ref{lagsd}) for Model II as
\ie
u_{2}(\lambda) = -4 \sqrt{1-\ell} \, M \ln (\lambda ),
\fe
so that
\ie
\mathfrak{p}_{\omega} =\int_0^\infty \left ( \alpha_{\omega\omega^\prime} \mathfrak{f}_{\omega^\prime} + \beta_{\omega\omega^\prime} \bar{ \mathfrak{f}}_{\omega^\prime}  \right)\mathrm{d} \omega^\prime,
\fe
with, for this case, 
\begin{equation}
\begin{split}
    \alpha_{\omega\omega^\prime}=& -i\mathfrak{K}e^{i\omega^\prime \mathfrak{v}_0}e^{\pi \left[2 M \sqrt{1-\ell} \right]\omega} \int_{-\infty}^{0} \,\mathrm{d}x\,\Big(\frac{\omega^\prime}{\omega}\Big)^{1/2}e^{\omega^\prime x} e^{i\omega\left[4M\sqrt{1-\ell}\right]\ln\left(\frac{|x|}{CD}\right)},
    \end{split}
\end{equation}
and
\begin{equation}
\begin{split}
    \beta_{\omega\omega'} &= i\mathfrak{K}e^{-i\omega^\prime \mathfrak{v}_0}e^{-\pi \left[2 M \sqrt{1-\ell} \right]\omega}
    \int_{-\infty}^{0} \,\mathrm{d}x\,\left(\frac{\omega^\prime}{\omega}\right)^{1/2}e^{\omega^\prime x}  e^{i\omega\left[4M\sqrt{1-\ell}\right]\ln\left(\frac{|x|}{CD}\right)},
    \end{split}
\end{equation}
After some algebraic manipulations, we obtain
\begin{equation}\label{eq:alphabetarel}
    |\alpha_{\omega\omega'}|^2 = e^{\big(8\pi M \sqrt{1-\ell} \big)\omega}|\beta_{\omega\omega'}|^2\,.
\end{equation}
Analogous to the previous section, we consider \(\omega\) to \(\omega + \mathrm{d}\omega\) as well, leading to
\ie
\label{Fomega}
    \mathfrak{P}_{2}(\omega, \ell)=\frac{\mathrm{d}\omega}{2\pi}\frac{1}{\left \lvert\frac{\alpha_{\omega\omega^\prime}}{\beta_{\omega\omega^\prime}}\right \rvert^2-1}\, ,
\fe
which follows
\ie
\label{power}
    \mathfrak{P}_{2}(\omega, \ell)=\frac{\mathrm{d}\omega}{2\pi}\frac{1}{e^{\left(8\pi M \sqrt{1-\ell}\right)\omega}-1}\,.
\fe
When the comparison with Planck distribution is taken into account, it reads
\begin{equation} \label{fffgg}
    \mathfrak{P}_{2}(\omega, \ell)=\frac{\mathrm{d}\omega}{2\pi}\frac{1}{e^{\frac{\omega}{T}}-1},
\end{equation}
which, finally, yields
\ie
\label{hawtemp}
    T_{2} = \frac{1}{8 \pi  \sqrt{1-\ell}\,  M}.
\fe

\begin{figure}
    \centering
      \includegraphics[scale=0.55]{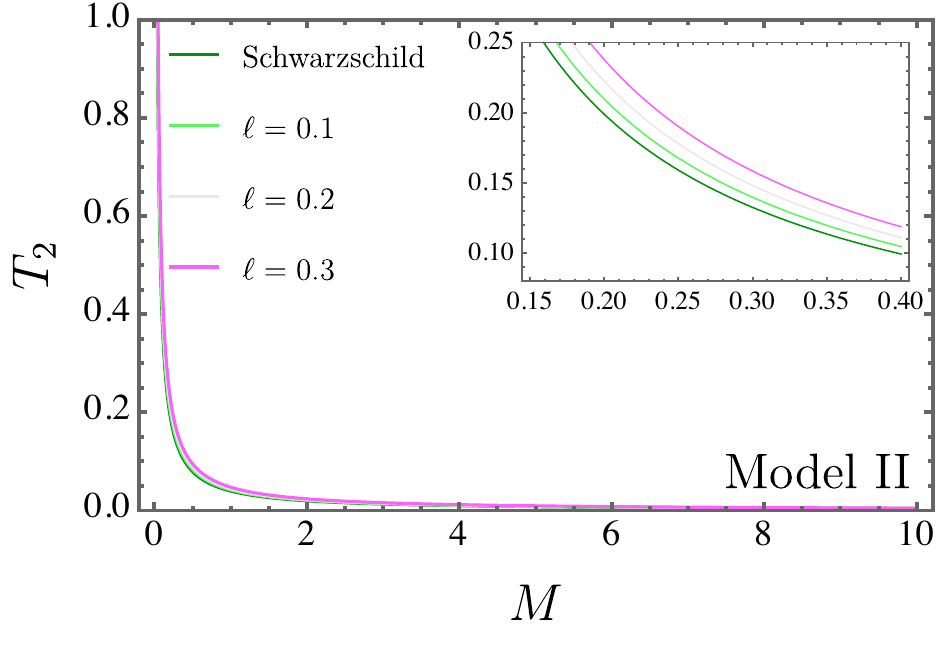}
    \caption{The \textit{Hawking} temperature $T_{2}$ as a function of mass \( M \) for various values of \( \ell \) in Model II.}
    \label{hawkingt2}
\end{figure}

As we will observe in the evaporation subsection, the \textit{Hawking} temperature found in Eq. (\ref{hawtemp}) aligns with that derived from surface gravity in Eq. (\ref{haah}), as expected. To provide a clearer understanding of this thermal quantity, its behavior is depicted in Fig. \ref{hawkingt2}, along with a comparison to the Schwarzschild case. Broadly, increasing \(\ell\) results in a rise in the \textit{Hawking} temperature. Additionally, consistent with the findings of the previous section, there is no remnant mass in this scenario.

As in Model I, since the black hole radiates, its mass steadily decreases, leading to its gradual reduction. In the following section, we shall apply the tunneling approach developed by Parikh and Wilczek \cite{011} to model this effect, similar to our treatment for Model I.

\subsubsection{The tunneling process}

\begin{figure}
    \centering
      \includegraphics[scale=0.55]{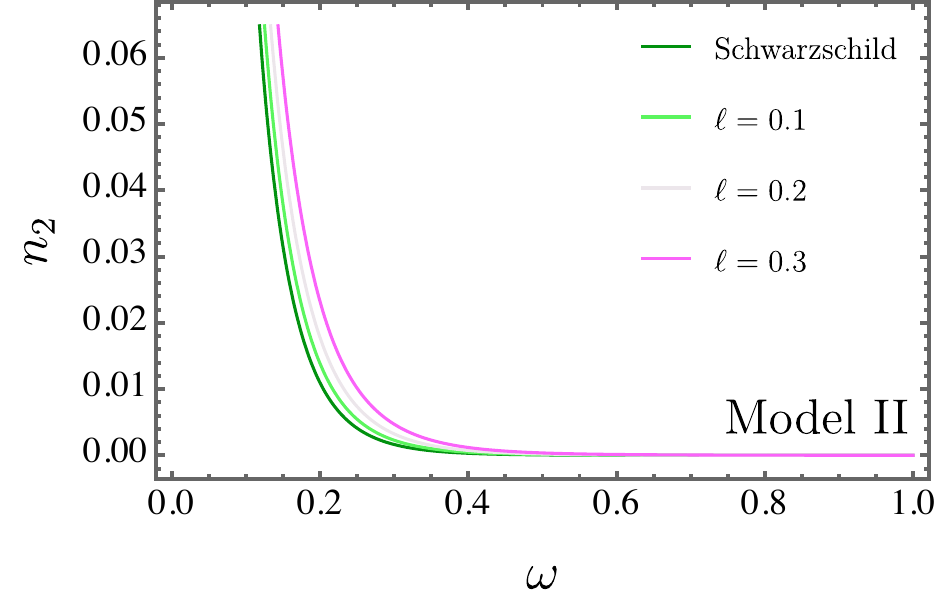}
    \caption{The particle density \( n_{2} \) is shown for different values of \( \ell \) in Model II.}
    \label{particledensitybosonstunneling2}
\end{figure}

Following the approach taken in the previous section for Model I, here we consider \(\Delta(r) = \frac{-\ell r + 2M }{1-\ell} \) so that the integral in Eq. (\ref{ims}) can be evaluated as shown below
\ie
\label{ims}
\text{Im}\, \mathcal{S} =\text{Im}\,\int_{0}^{\omega} -\mathrm{d}\omega'\int_{r_i}^{r_f}\,\frac{\mathrm{d}r}{ \sqrt{1-\ell}\left( 1-\sqrt{\frac{\Delta(r)}{r}}\right)}.
\fe
The function \(\Delta(r)\) will depend on \(\omega'\), resulting from substituting \(M\) with \((M - \omega')\) in the metric. This modification creates a pole at the new horizon location \(r = \mathfrak{r}\). Integrating along a counterclockwise path around this pole produces
\begin{eqnarray}
    \text{Im}\, \mathcal{S}  = 4\pi \sqrt{1-\ell} \, \omega \left( M - \frac{\omega}{2} \right)  .
\end{eqnarray}
Thus, the Lorentz--violating correction to the emission rate for a \textit{Hawking} particle is given by 
\ie
\Gamma_{2} \sim e^{-2 \, \text{Im}\, S}=e^{-8 \pi \sqrt{1-\ell} \, \omega \left( M - \frac{\omega}{2} \right)} .
\fe
Notably, in the limit \(E \to 0\), the standard Planckian spectrum as initially derived by Hawking is restored. Consequently, the emission spectrum is given by
\begin{equation}
    \mathfrak{P}_{2}(\omega)=\frac{\mathrm{d}\omega}{2\pi}\frac{1}{e^{8 \pi \sqrt{1-\ell} \, \omega \left( M - \frac{\omega}{2} \right)
    }-1}.
\end{equation}
With its additional dependence on \(\omega\), the emission spectrum departs from the classic black body distribution, as can be readily observed. For small values of \(\omega\), the expression reduces to the Planck distribution but with an adjusted \textit{Hawking} temperature. Furthermore, the particle number density can be determined using the tunneling rate
\ie
n_{2} = \frac{\Gamma_{2}}{1 - \Gamma_{2}} = \frac{1}{e^{8 \pi  \sqrt{1-\ell} \, \omega  \left(M-\frac{\omega }{2}\right)} - 1}
\fe
To provide a clearer understanding of \(n_{2}\), we present Fig. \ref{particledensitybosonstunneling2}, which illustrates its dependence on the Lorentz-violating parameter \(\ell\). As \(\ell\) increases, we observe an increase in particle number density. Additionally, \(n_{2}\) is compared with the Schwarzschild case. In contrast to \(n_{1}\), the lines representing \(n_{2}\) are spaced more closely together.

\begin{figure}
    \centering
      \includegraphics[scale=0.55]{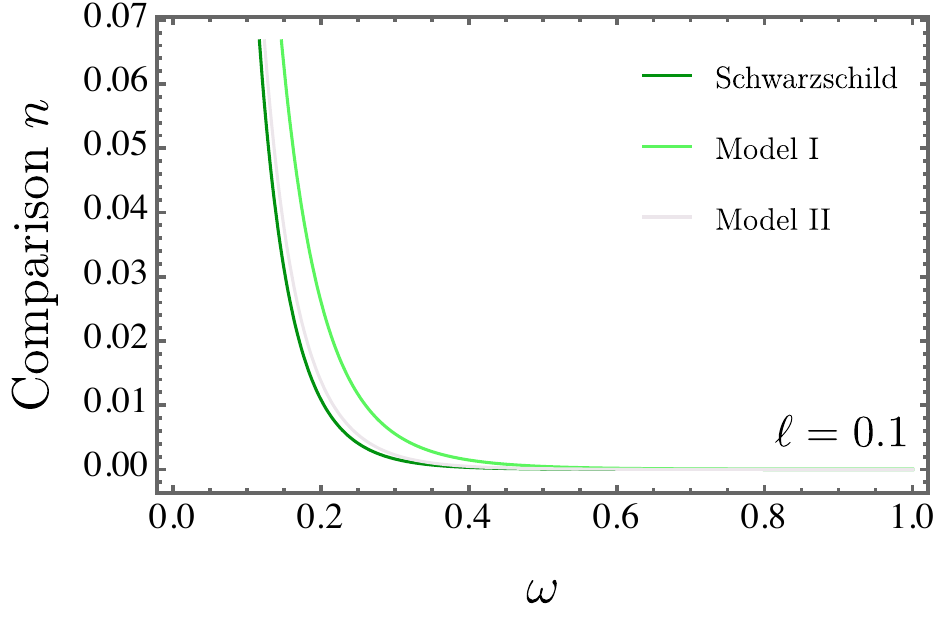}
    \caption{Comparison of \( n \) for Model I and Model II with \( \ell = 0.1 \) and \( M = 1 \), alongside the Schwarzschild case.}
    \label{compppp}
\end{figure}

We further compare the particle density for both models developed here in the bosonic case, denoted by \(n_{1}\) and \(n_{2}\), in Fig. \ref{compppp}. The Schwarzschild case is also included for reference. In a general panorama, the particle density intensities exhibit the following hierarchy: \(n\) (Model I) \(> n_{2}\) (Model II) \(> \) Schwarzschild.

\subsection{Fermionic modes}

\begin{figure}
    \centering
      \includegraphics[scale=0.55]{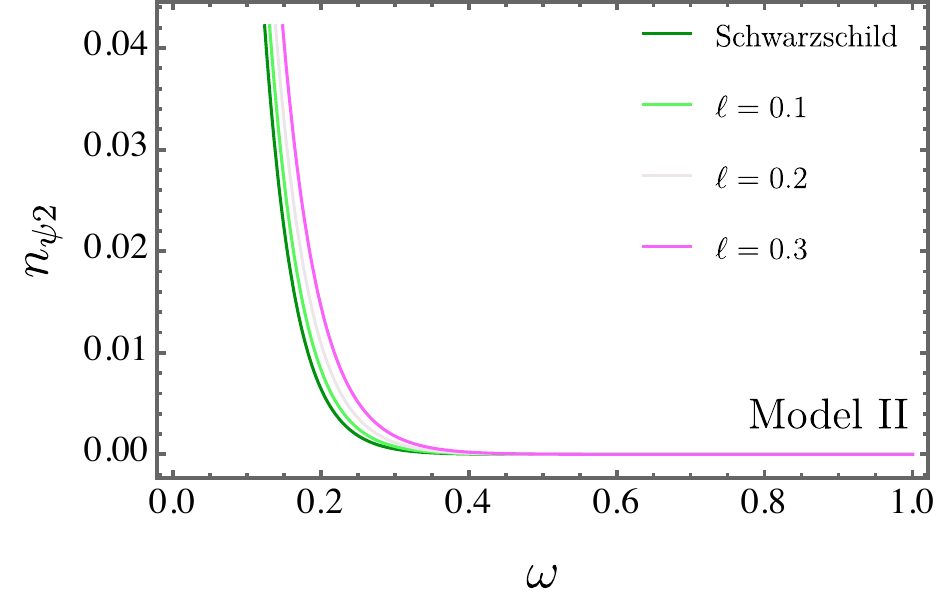}
    \caption{The particle density \( n_{\psi 2} \) is shown for different values of \( \ell \) in Model II.}
    \label{particlesfermionsmodel2}
\end{figure}

\begin{figure}
    \centering
      \includegraphics[scale=0.55]{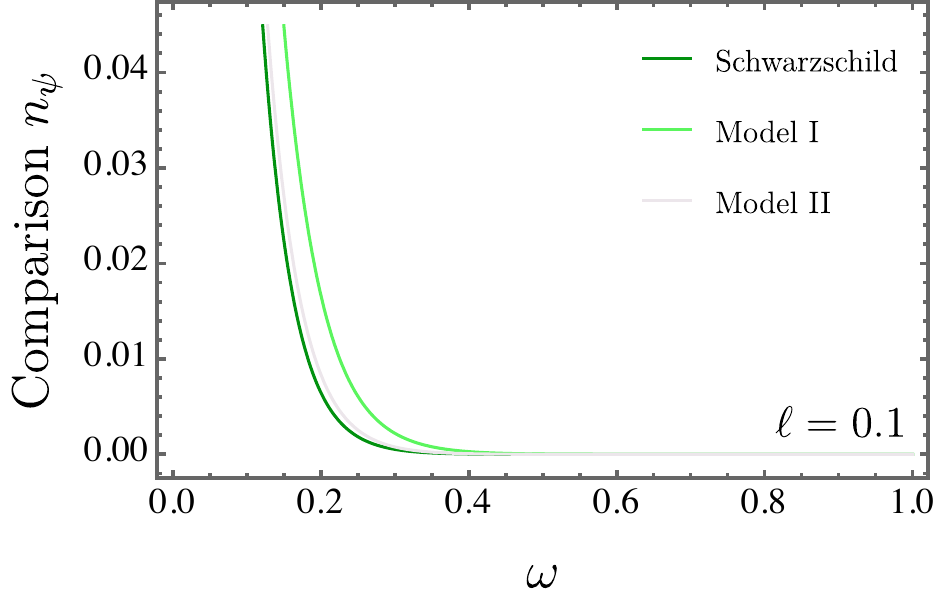}
    \caption{Comparison of \( n_{\psi} \) for Model I and Model II with \( \ell = 0.1 \) and \( M = 1 \), alongside the Schwarzschild case.}
    \label{comparionsonfermions}
\end{figure}

Using all definitions outlined thus far, we can now express the particle density for fermions in Model II as follows:
\ie
n_{\psi2} = \frac{\Gamma_{\psi2}}{1+\Gamma_{\psi2}}  = \frac{1}{e^{\frac{8 \pi  \omega }{\sqrt{\frac{1}{M^2-\ell M^2}}}}+1}.
\fe
In Fig. \ref{particlesfermionsmodel2}, we illustrate the behavior of \(n_{\psi2}\) across different values of \(\ell\) and compare these findings with the standard Schwarzschild case. Additionally, Fig. \ref{comparionsonfermions} compares both models with the Schwarzschild case. Overall, we observe that Model I exhibits a higher particle density than Model II. It is also important to note that the Schwarzschild case shows the lowest particle density among them.

\subsection{Greybody factors for bosons}

\begin{figure}
    \centering
      \includegraphics[scale=0.55]{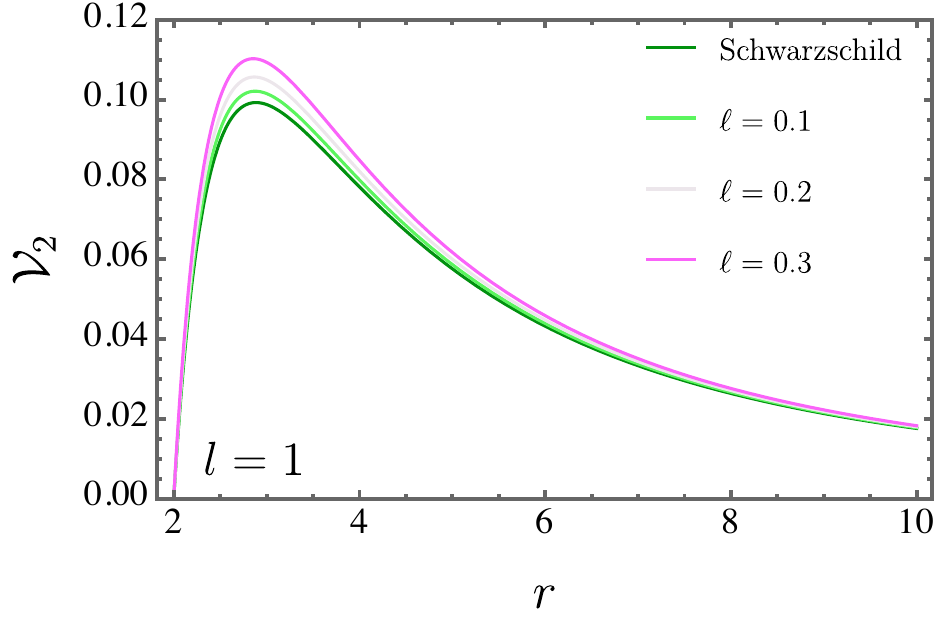}
    \caption{The effective potential for bosons $\mathcal{V}_{2}$ is shown for different values of the Lorentz--violating parameter $\ell$ with a fixed angular momentum quantum number $l=1$. The Schwarzschild case is also included for comparison in this analysis.}
    \label{potentialbosonmodelII}
\end{figure}

Following the approach taken in the previous section for Model I, a similar analysis is conducted here for Model II. In this context, the term $\mathcal{V}_{2}$ describes the effective potential governing scalar perturbations, given by
\ie
\begin{split}
\mathcal{V}_{2} & = \mathcal{A}(r)\left[\frac{{l(l + 1)}}{{{r^2}}} + \frac{1}{{r\sqrt {{\mathcal{A}(r)}{\mathcal{B}(r)^{ - 1}}} }}\frac{\mathrm{d}}{{\mathrm{d}r}}\sqrt {{\mathcal{A}(r)}\mathcal{B}(r)}\right]\\
& = \frac{l(l+1)(r - 2M) + \frac{2 M \sqrt{-\frac{(r-2 M)^2}{(\ell - 1) r^2}}}{\sqrt{1 - \ell}}}{r^3}.
\end{split}
\fe
Fig. \ref{potentialbosonmodelII} displays the effective potential $\mathcal{V}_{2}$ for scalar perturbations. In contrast to Model I, the potential increases as the parameter $\ell$ grows. The Schwarzschild case, included for reference, exhibits the lowest magnitude.

Additionally, the greybody factors for Model II are expressed as
\ie
\begin{split}
& T_{b2}  \ge {\mathop{\rm sech}\nolimits} ^2 \left[\int_{-\infty}^ {+\infty} \frac{\mathcal{V}_{2}} {2\omega}\mathrm{d}r^{*}\right] \\
& ={\mathop{\rm sech}\nolimits} ^2 \left[\int_{r_{ h}}^ {+\infty} \frac{\mathcal{V}_{2}} {2\omega\sqrt{\mathcal{A}(r)\mathcal{B}(r)}}\mathrm{d} r\right] \\
& ={\mathop{\rm sech}\nolimits} ^2 \left[ \frac{1}{2\omega} \left(\frac{ 1 + 2 l (l+1) (1 - \ell) }{4 \sqrt{ 1- \ell} \, M} \right)  \right].
\end{split}
\fe

\begin{figure}
    \centering
      \includegraphics[scale=0.51]{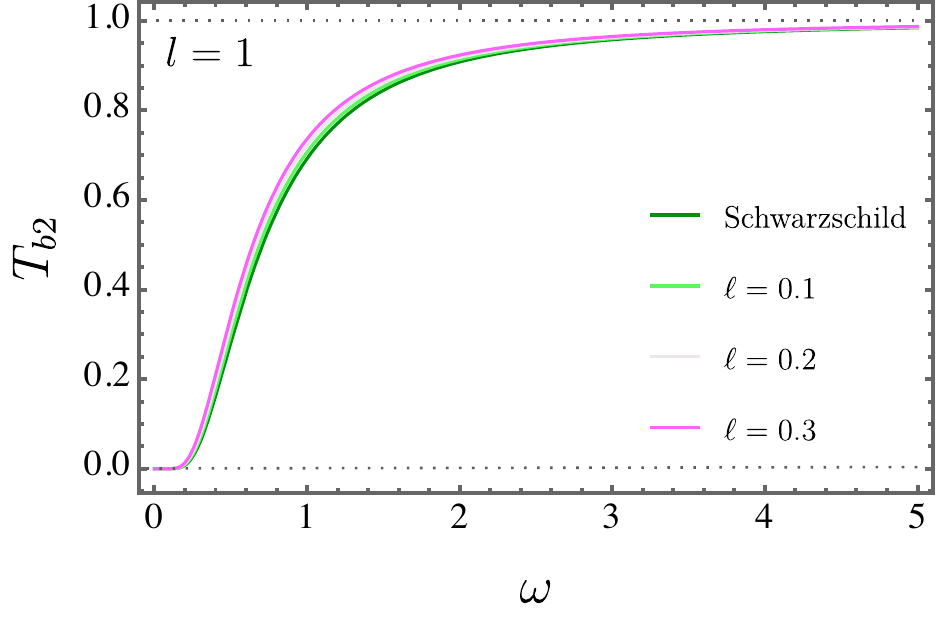}
        \includegraphics[scale=0.51]{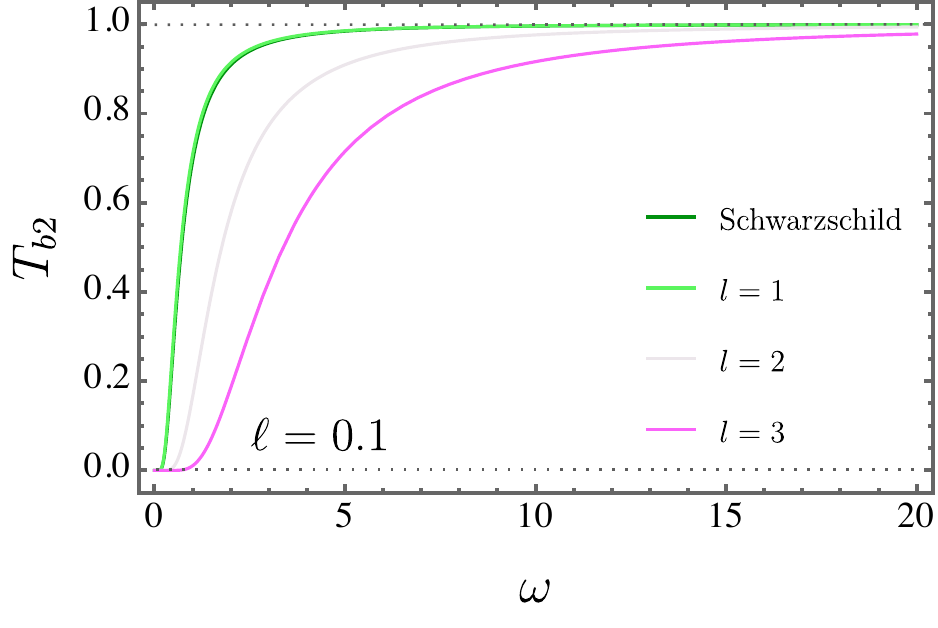}
    \caption{The greybody factors for bosons of Model II, denoted as $T_{b2}$, are presented for varying values of Lorentz--violating parameter $\ell$ with a fixed angular momentum quantum number $l=1$ (left panel), and for varying values of $l$ with a fixed $\ell$ (right panel). The Schwarzschild case is included for comparison in this analysis.}
    \label{greybodybosonsmodelII}
\end{figure}

Fig. \ref{greybodybosonsmodelII} shows the greybody factors for bosons, $T_{b2}$, under two conditions: the left panel illustrates variations in $\ell$ with $l=1$ fixed, while the right panel displays changes in $l$ with $\ell$ kept constant at 0.1. Both panels include the Schwarzschild case for comparison. Unlike Model I, the greybody factors here increase in magnitude as $\ell$ grows, with the Schwarzschild black hole exhibiting the lowest values.

\begin{figure}
    \centering
      \includegraphics[scale=0.55]{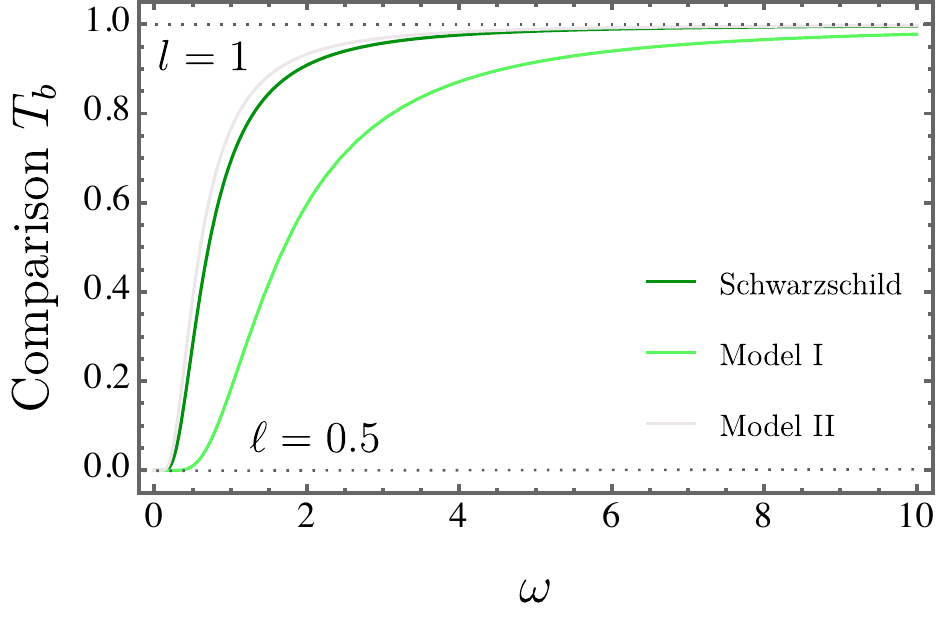}
    \caption{Comparison of $T_{b}$ for Model I and Model II with \( \ell = 0.5 \) and \( M = 1 \), alongside the Schwarzschild case.}
    \label{comppppgrebodybosons}
\end{figure}

Having calculated the greybody factors for bosons in both models, Fig. \ref{comppppgrebodybosons} provides a comparative analysis. Overall, Model II exhibits the highest magnitude, while Model I shows the lowest, even when compared to the Schwarzschild black hole.

\subsection{Greybody factors for fermions}

Similar to the approach taken in the previous section for the fermionic case in Model I, the term $\mathcal{V}_{2\psi}$ here describes the effective potential governing scalar perturbations, given by
\ie
\begin{split}
\mathcal{V}^{+}_{2\psi} & = \frac{(l + \frac{1}{2})^2}{r^2} \mathcal{A}(r)  \pm \left(l + \frac{1}{2}\right) \sqrt{\mathcal{A}(r) \mathcal{B}(r)} \partial_r \left(\frac{\sqrt{\mathcal{A}(r)}}{r}\right)\\
& = \frac{(2 l+1) (3 M-r) \sqrt{\frac{(r-2 M)^2}{(1 - \ell) r^2}}}{2 r^3 \sqrt{1-\frac{2 M}{r}}}+\frac{\left(l+\frac{1}{2}\right)^2 \left(1-\frac{2 M}{r}\right)}{r^2}.
\end{split}
\fe

\begin{figure}
    \centering
      \includegraphics[scale=0.55]{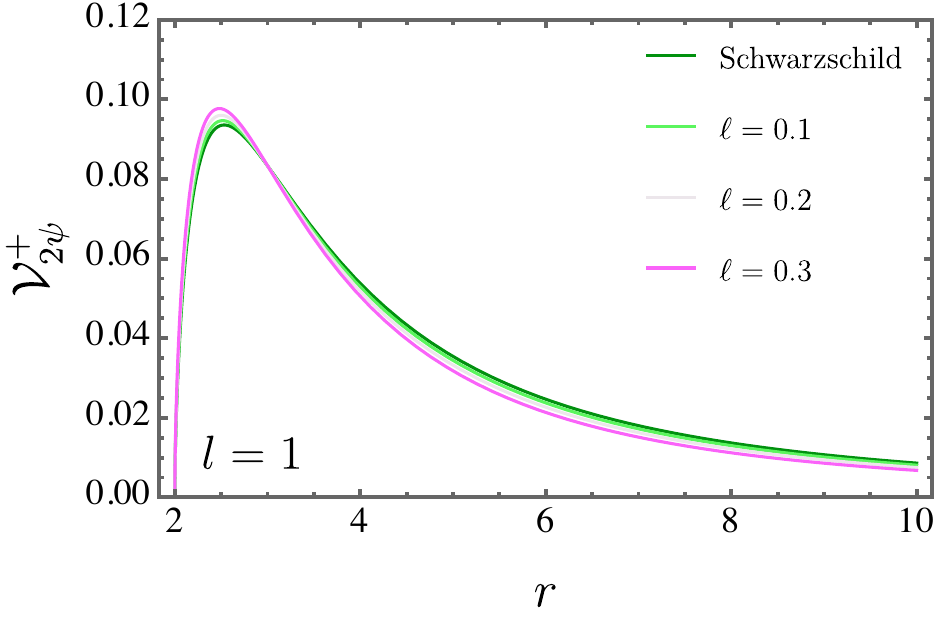}
    \caption{The effective potential for fermions $\mathcal{V}^{+}_{2\psi}$ is shown for different values of the Lorentz--violating parameter $\ell$ with a fixed angular momentum quantum number $l=1$. The Schwarzschild case is also included for comparison in this analysis.}
    \label{potentialfermionsmodelII}
\end{figure}

Fig. \ref{potentialfermionsmodelII} depicts the behavior of the effective potential $\mathcal{V}^{+}_{2\psi}$, being compared to the Schwarzschild case. Finally, the greybody factors for fermions in Model II are given by
\ie
\begin{split}
& T_{b2\psi}  \ge {\mathop{\rm sech}\nolimits} ^2 \left[\int_{-\infty}^ {+\infty} \frac{\mathcal{V}^{+}_{2\psi}} {2\omega}\mathrm{d}r^{*}\right] \\
& ={\mathop{\rm sech}\nolimits} ^2 \left[\int_{r_{ h}}^ {+\infty} \frac{\mathcal{V}^{+}_{2\psi}} {2\omega\sqrt{\mathcal{A}(r)\mathcal{B}(r)}}\mathrm{d} r\right] \\
& ={\mathop{\rm sech}\nolimits} ^2 \left[ \frac{1}{2\omega} \left(\frac{(2 l+1)^2 \sqrt{1 - \ell}}{8 M} \right)  \right].
\end{split}
\fe

\begin{figure}
    \centering
      \includegraphics[scale=0.51]{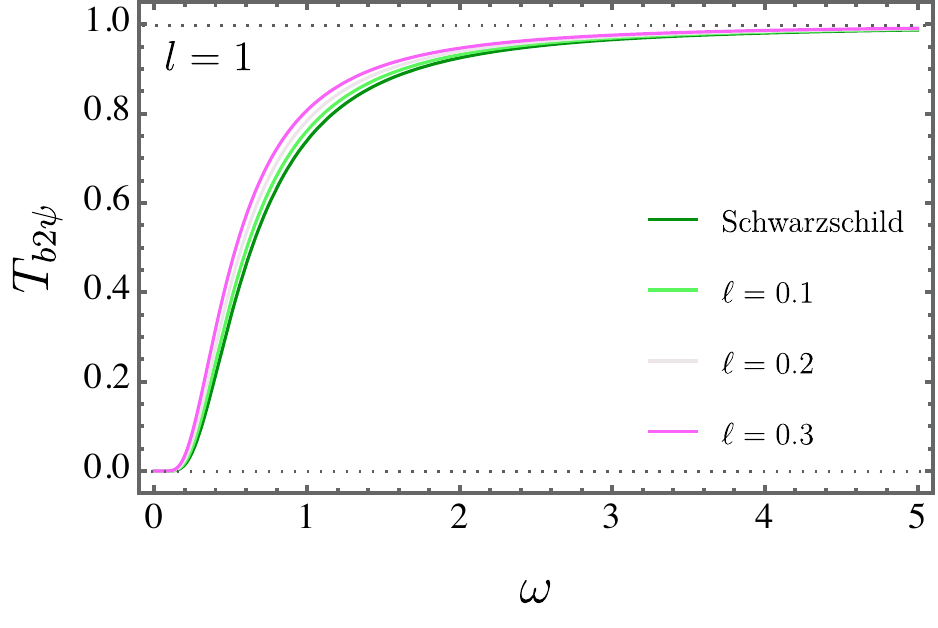}
        \includegraphics[scale=0.51]{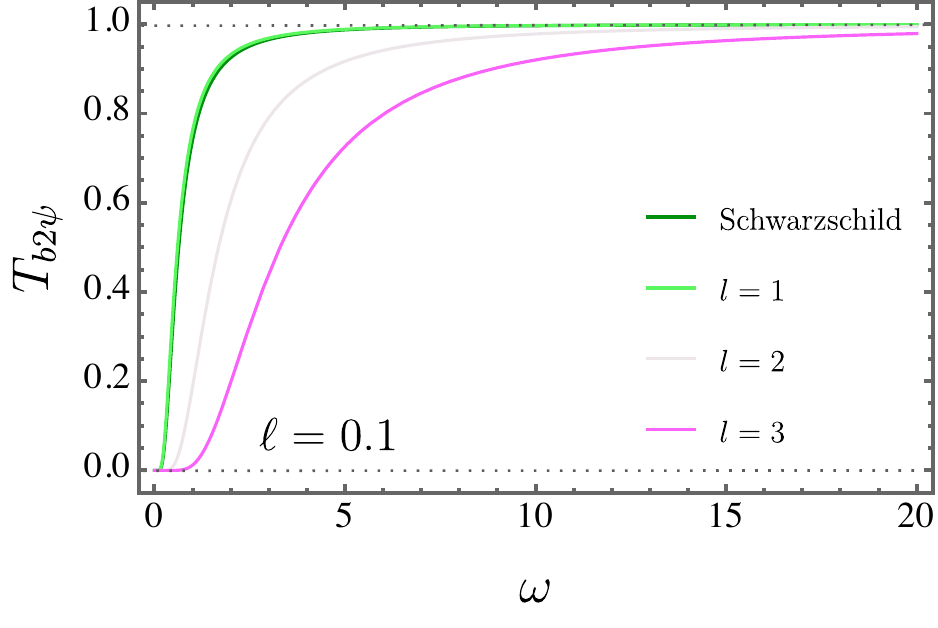}
    \caption{The greybody factors for fermions of Model II, denoted as $T_{b2\psi}$, are presented for varying values of Lorentz--violating parameter $\ell$ with a fixed angular momentum quantum number $l=1$ (left panel), and for varying values of $l$ with a fixed $\ell$ (right panel). The Schwarzschild case is included for comparison in this analysis.}
    \label{greybodyfermionsmodelII}
\end{figure}

Fig. \ref{greybodyfermionsmodelII} shows the greybody factors for fermions, $T_{b2\psi}$, in two cases: the left panel explores variations in $\ell$ with $l=1$ fixed, while the right panel examines changes in $l$ with $\ell$ set to $0.1$. Both panels include the Schwarzschild case for comparison.

\begin{figure}
    \centering
      \includegraphics[scale=0.55]{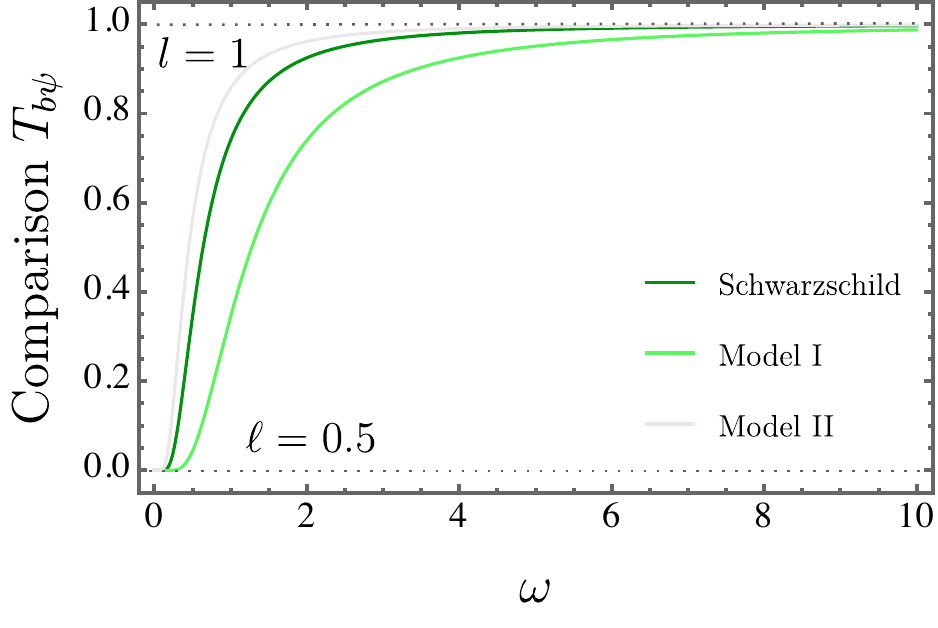}
    \caption{Comparison of $T_{b\psi}$ for Model I and Model II with \( \ell = 0.5 \) and \( M = 1 \), alongside the Schwarzschild case.}
    \label{comppppgrebodyfermions}
\end{figure}

After computing the greybody factors for fermions in both models, Fig. \ref{comppppgrebodyfermions} compares their behavior. For the configuration with $l=1$ and $\ell=0.5$, Model II exhibits the highest magnitude.

\begin{figure}
    \centering
      \includegraphics[scale=0.55]{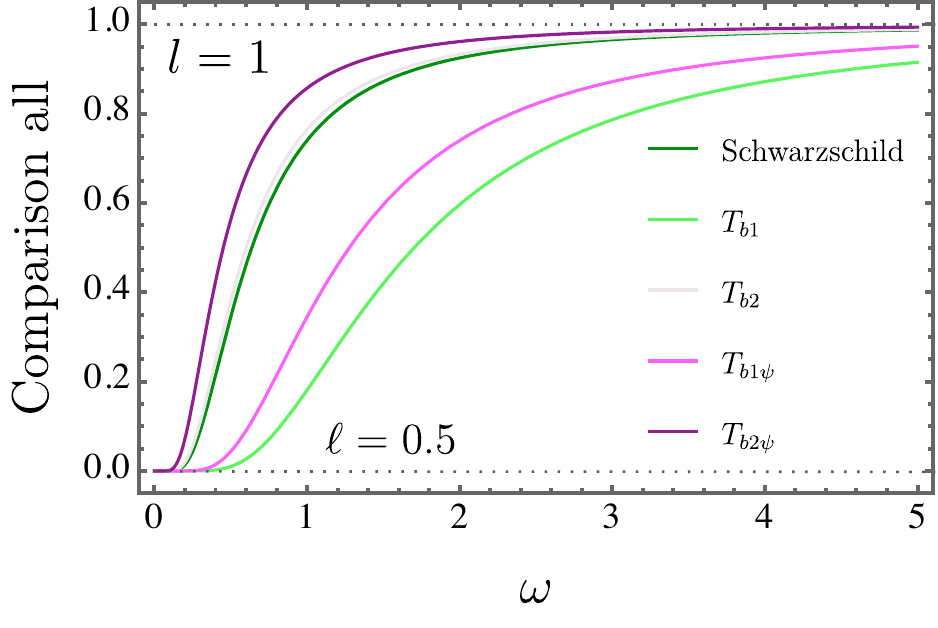}
    \caption{All greybody factors for Model I and Model II with $\ell = 0.5$ and $M = 1$ are compared against the Schwarzschild case.}
    \label{compall}
\end{figure}

To ensure a comprehensive comparison, one might ask whether the greybody factors exhibit higher magnitudes when considering all spins (bosons and fermions) for both models. Fig. \ref{compall} addresses this question, showing that Model II (for both bosons and fermions) exhibits greater magnitudes compared to Model I and Schwarzschild case. The greybody factors for fermions are consistently larger than those for bosons within each respective model. Moreover, Model I (for both bosons and fermions) presents lower magnitudes compared to the Schwarzschild case.

\subsection{The evaporation process}

In Model II, the Lorentz violation parameter \(\ell\) has no effect on the event horizon, photon sphere, or shadow radii. However, it does modify the \textit{Hawking} temperature, leading to a difference in the evaporation process when compared to the Schwarzschild case. Therefore, applying the surface gravity method and adopting a similar approach to the calculations performed for Model I, as shown in Eqs. (\ref{surface1}), (\ref{surface2}), and (\ref{surface3}), we obtain \cite{kanzi2019gup}
\ie
\nonumber
T_{2} = \frac{1}{8 \pi  \sqrt{1-\ell} \, M},
\fe
where, once again, the Hawking temperature is consistent with the result obtained through the analysis of Hawking radiation via the Bogoliubov coefficients, as expected. Notice that, in Model II, the event horizon $\mathfrak{r}^{(\mathrm{II})}$ coincides with the Schwarzschild event horizon, i.e., $\mathfrak{r}^{(\mathrm{II})} = 2M$.

\begin{figure}
    \centering
      \includegraphics[scale=0.55]{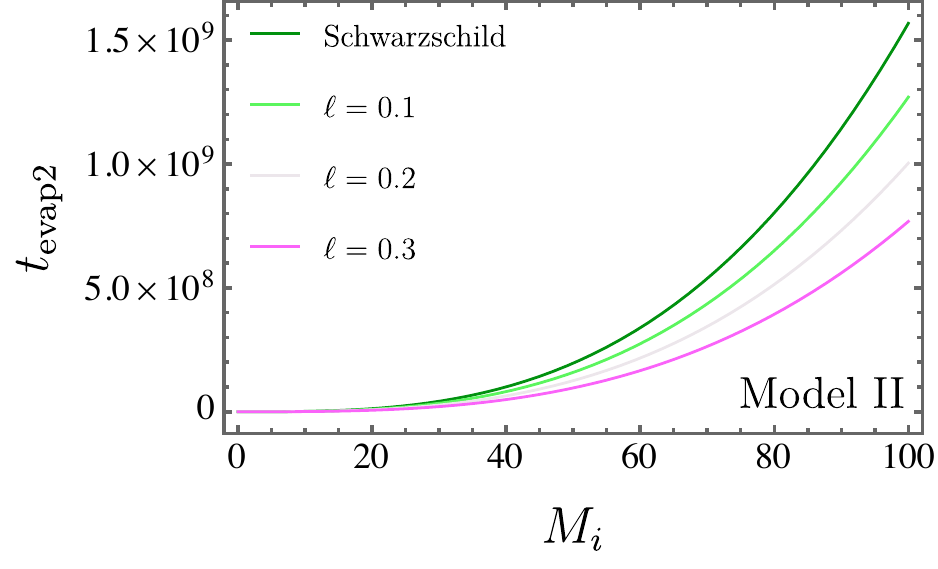}
    \caption{The evaporation time \( t_{\text{evap}} \) is shown for different values of \( \ell \).}
    \label{evap2}
\end{figure}

\begin{figure}
    \centering
      \includegraphics[scale=0.55]{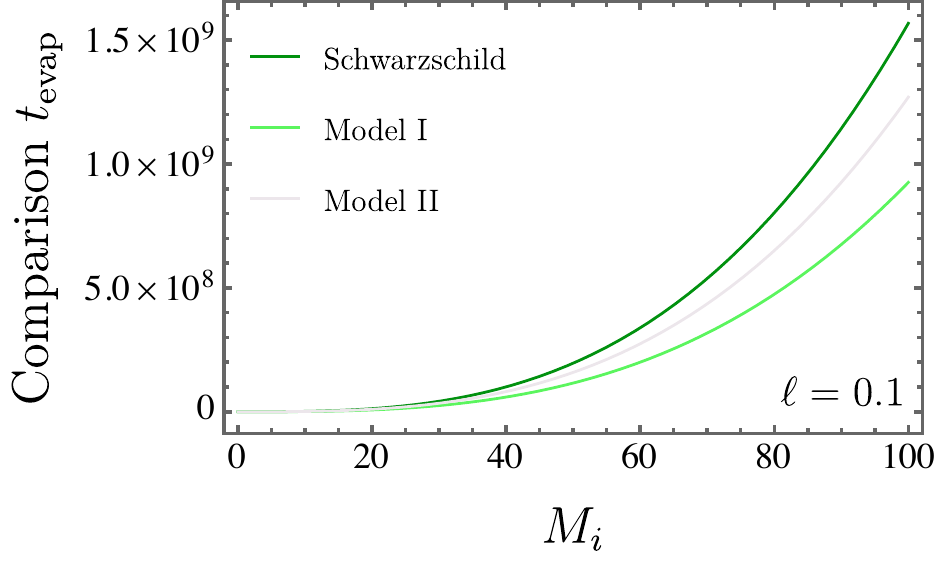}
    \caption{Comparison of the evaporation time \( t_{\text{evap}}  \) for Model I and Model II when a fixed value of \( \ell = 0.1 \) is considered.}
    \label{comparisont}
\end{figure}

As in the previous section, we will apply the \textit{Stefan--Boltzmann} law from Eq. (\ref{sflaw}) to analyze the black hole evaporation process. Here, we set \(\sigma = \pi (3\sqrt{3}M)^{2}\), yielding
\ie
\begin{split}
\int_{0}^{t_{\text{evap2}}} \xi \mathrm{d}\tau & = - \int_{M_{i}}^{M_{f}} 
\left[ \frac{27 \xi}{4096 \pi ^3 (1-\ell)^2 M^2}  \right]^{-1} \mathrm{d}M.
\end{split}
\fe
Accordingly, it is given by
\ie
t_{\text{evap2}} = - \frac{4096 \pi ^3 (1-\ell)^2 \left(M_{f}^3 - M_{i}^3\right)}{81 \xi}.
\fe
For this black hole model, analogous to the Model I, no remnant mass is expected; thus, we assume it will evaporate completely, meaning \(M_{f} \to 0\).
\ie
t_{\text{evap2}} = \frac{4096 \pi ^3 (1-\ell)^2 M_{i}^3}{81 \xi}.
\fe
To clarify our results, we plot \(t_{\text{evap2}}\) in Fig. \ref{evap2} for various values of \(\ell\). As \(\ell\) increases, evaporation time decreases. For a comparative analysis of the two models developed thus far, we present
\ie
\frac{t_{\text{evap}}}{t_{\text{evap2}}} = (1 - \ell)^{3}.
\fe
This indicates that the evaporation time for Model I is shorter than for Model II, since \(\ell \ll 0\) \cite{Junior:2024ety}. Consequently, Model I undergoes evaporation more rapidly than Model II. To illustrate this difference, we plot both models at a fixed value of \(\ell = 0.1\) to directly compare their evaporation times. This is shown in Fig. \ref{comparisont}. Overall, both Models I and II exhibit faster evaporation rates than the Schwarzschild case. However, between the two, Model I consistently evaporates faster than Model II.

In our results derived throughout this paper, we observe that, at least for the purpose of visualizing the distinct influence of $\ell$ on black hole properties in Kalb--Ramond gravity compared to the Schwarzschild case, $\ell$ is practically suppressed due to its extremely small magnitude \cite{yang2023static}. For this reason, we selected the representative values $\ell = 0.1, 0.2$, and $0.3$ to enhance visualization of the deviations from the Schwarzschild case, following a similar approach adopted in recent studies on Lorentz violation within the framework of bumblebee gravity \cite{filho2023vacuum,AraujoFilho:2024ykw,heidari2024scattering}. To overcome this limitation and provide greater robustness to the interpretations, the next section presents the dimensionless ratios of the quantities of interest (Hawking temperature, evaporation lifetime, and particle creation density) as functions of the parameter $\ell$, in accordance with the bounds established in the literature \cite{yang2023static}.

 \section{Deviation from general relativity}

In this section, we present an investigation of the departure of general relativity ascribed to the Lorentz violation, which, in our case, is represented by $\ell$. Therefore, we shall use these bounds as a guide for the permissible range of $\ell$ values. Let $T_H^{(\mathrm{GR})}$, $t_{\text{evap}}^{(\mathrm{GR})}$, and $n^{(\mathrm{GR})}$ represent the standard GR quantities, namely, the Hawking temperature, evaporation time, and particle creation density, respectively. In particular, focusing on the Hawking temperature in the Models analyzed here, the KR modification introduces factors of $1/(1-\ell)^{\Bar{\alpha}}$ (where $\Bar{\alpha}$ depends on the Model details naturally). As we have derived in the previous sections,
\ie
   T_{H}^{(\mathrm{I})}
   \;=\;
    \frac{1}{8 \pi  (1- \ell)^2 M}
   \quad \text{(Model I)},
   \quad
   T_{H}^{(\mathrm{II})}
   \;=\; 
    \frac{1}{8 \pi  \sqrt{1-\ell}\,  M}
   \quad \text{(Model II)}.
\fe
Notice that, by writing
\ie
   T_H^{(\mathrm{KR})} \;=\; T_H^{(\mathrm{GR})}\, \frac{1}{(1-\ell)^{\Bar{\alpha}}},
\fe
then the relative deviation from GR is
\ie
   \frac{T^{(\mathrm{KR})}_{H}}{T_H^{(\mathrm{GR})}}
   \;=\;
    \frac{1}{(1-\ell)^{\Bar{\alpha}}} ,
\fe
and, for small $\ell$, we can expand $1/(1-\ell)^{\Bar{\alpha}} \approx 1 + \Bar{\alpha}\,\ell$. In this manner,
\ie
\bar{T} =   \frac{ T_{H}^{(\mathrm{KR})}}{T_{H}^{(\mathrm{GR})}}
   \;\approx\; 1 +\, \Bar{\alpha}\,\ell,
   \quad
   \text{with }
   \Bar{\alpha} = 
   \begin{cases}
     2,\;\text{(Model I)},\\
     1/2,\;\text{(Model II)}.
   \end{cases}
\fe

\begin{figure}
    \centering
      \includegraphics[scale=0.45]{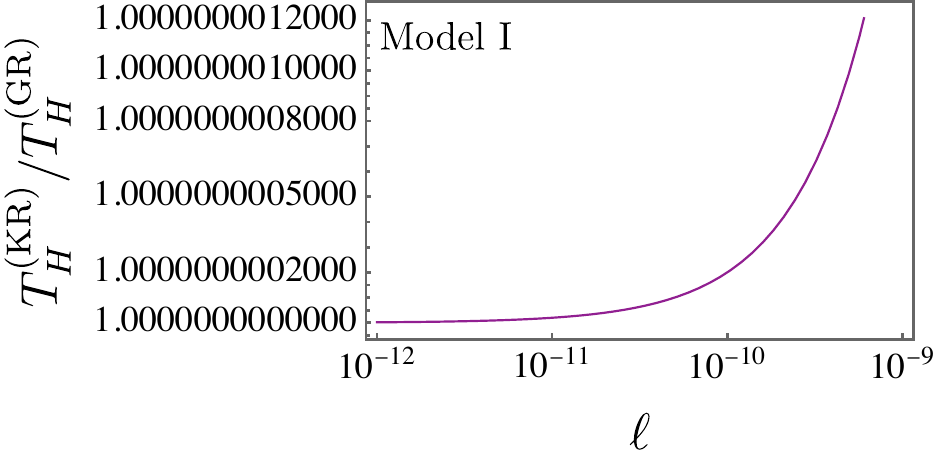}
       \includegraphics[scale=0.45]{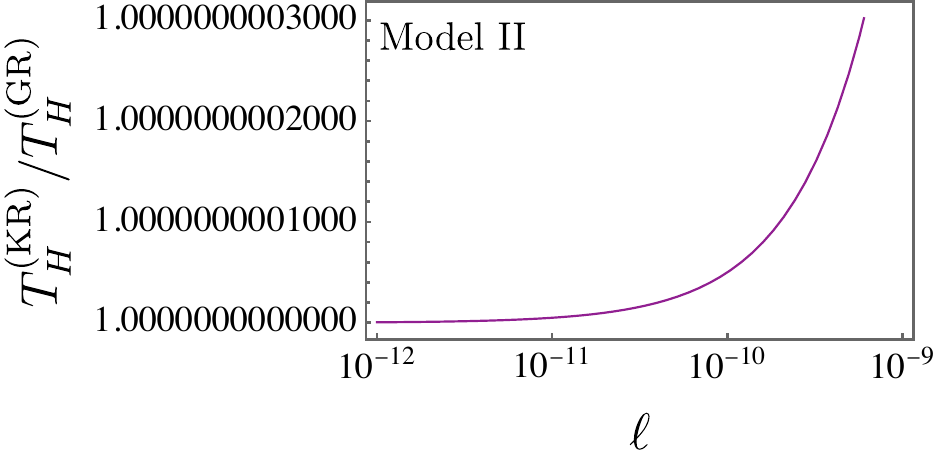}
    \caption{ The ratio $T_H^{(\mathrm{KR})}/T_H^{(\mathrm{GR})}$ is plotted as a function of $\ell$ for both models, using a logarithmic scale on both axes. }
    \label{depaturefromgrt}
\end{figure}

In Fig. \ref{depaturefromgrt}, we illustrate how the Hawking temperature deviates in Kalb--Ramond gravity for Model I and Model II, compared to general relativity. In other words, the variation of $T_H^{(\mathrm{KR})}/T_H^{(\mathrm{GR})}$ is plotted as a function of $\ell$. Analogous to this latter case, for improved visualization, all subsequent plots will adopt a logarithmic scale on both axes (“log-log” plots), with the range of $\ell$ chosen in accordance with the bounds recently discussed in the literature \cite{yang2023static}.

\begin{figure}
    \centering
      \includegraphics[scale=0.45]{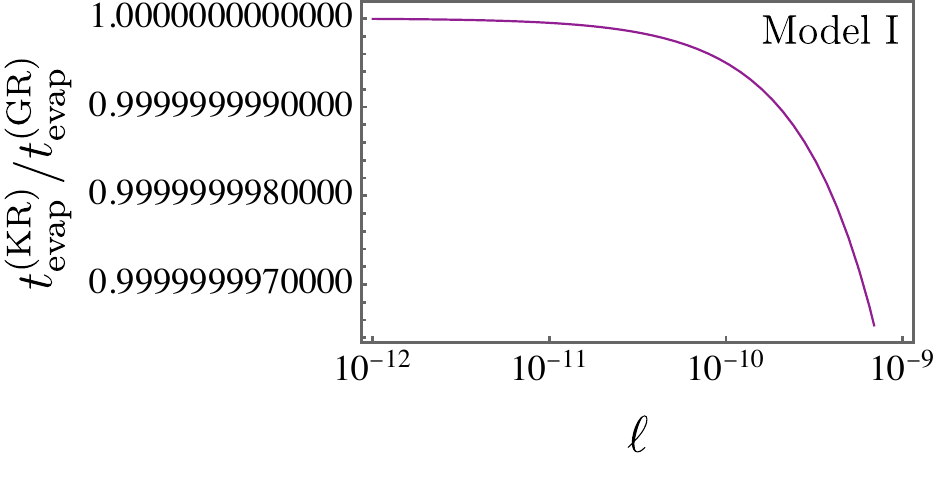}
       \includegraphics[scale=0.45]{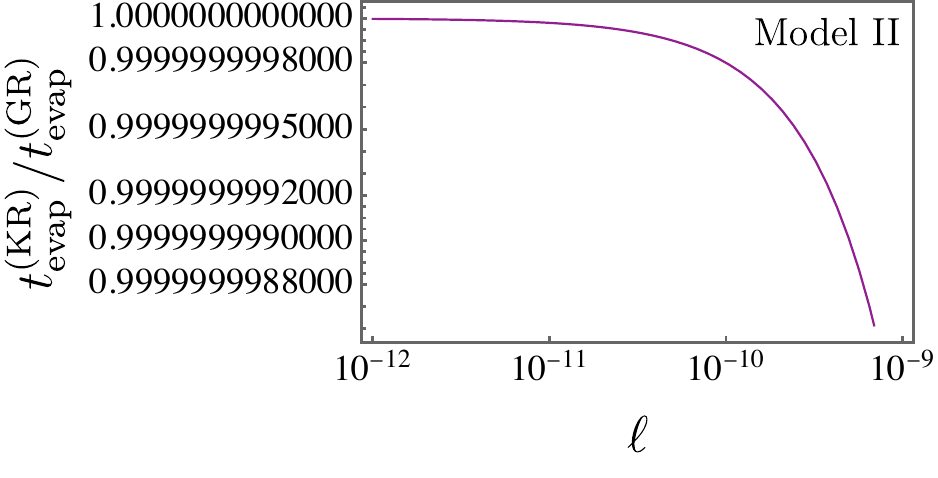}
    \caption{ The ratio $t_{\text{evap}}^{(\mathrm{KR})}/t_{\text{evap}}^{(\mathrm{GR})}$ is plotted as a function of $\ell$ for both models, employing a logarithmic scale on both axes. }
    \label{depaturefromgrevap}
\end{figure}

In addition, considering the evaporation lifetime comparison, we have:
\ie
   t_{\text{evap}}^{(\mathrm{I})}
   \;=\;
   \frac{4096 \,\pi^3\, (1-\ell)^5\,M_i^3}{81\,\xi}
   \quad \text{(Model I)},
   \quad
   t_{\text{evap}}^{(\mathrm{II})}
   \;=\;
   \frac{4096 \,\pi^3\, (1-\ell)^2\,M_i^3}{81\,\xi}
   \quad \text{(Model II)}.
\fe
Comparing to the standard GR scaling $t_{\text{evap}}^{(\mathrm{GR})} \propto M_i^3$, we see that
\ie
   t_{\text{evap}}^{(\mathrm{KR})}
   \;=\;
   t_{\text{evap}}^{(\mathrm{GR})}\,\bigl[(1-\ell)^{\Bar{\beta}}\bigr].
\fe
If $\ell \ll 1$, then $(1-\ell)^{\Bar{\beta}} \approx 1 - {\Bar{\beta}} \,\ell$, we have 
\ie
\bar{t} =   \frac{ t^{(\mathrm{KR})}_{\text{evap}}}{t_{\text{evap}}^{(\mathrm{GR})}}
   \;\approx\; 1 -\,{\Bar{\beta}}\,\ell,
   \quad
   \text{with }
   {\Bar{\beta}} = 
   \begin{cases}
     5,\;\text{(Model I)},\\
     2,\;\text{(Model II)}.
   \end{cases}
\fe
In Fig. \ref{depaturefromgrevap}, the behavior of the ratio $t_{\text{evap}}^{(\mathrm{KR})}/t_{\text{evap}}^{(\mathrm{GR})}$ as a function of $\ell$ is shown for both Models
Next, we examine the deviation from general relativity in the particle creation density. As seen in the previous sections, we have obtained
\ie
   n^{(\mathrm{I})}
   \;=\;
    \frac{1}{e^{8 \pi  (1-\ell)^2 \omega  \left(M-\frac{\omega }{2}\right)} - 1}
   \quad \text{(Model I)},
   \quad
   n^{(\mathrm{II})}
   \;=\;
    \frac{1}{e^{8 \pi  \sqrt{1-\ell} \, \omega  \left(M-\frac{\omega }{2}\right)} - 1}
   \quad \text{(Model II)}.
\fe
In this case, notice that 
\ie
   \Gamma^{(\mathrm{I})}
   \;=\;
    e^{8 \pi  (1-\ell)^2 \omega  \left(M-\frac{\omega }{2}\right)}
   \quad \text{(Model I)},
   \quad
   \Gamma^{(\mathrm{II})}
   \;=\;
    e^{8 \pi  \sqrt{1-\ell} \, \omega  \left(M-\frac{\omega }{2}\right)}
   \quad \text{(Model II)},
\fe
so that the following relation is straightforwardly derived
\ie
   \Gamma^{(\mathrm{KR})}
   \;=\;
   \Gamma^{(\mathrm{GR})}\,e^{8 \pi  \omega  \left[(1-l)^{\Bar{\alpha} }-1\right] \left( M - \frac{\omega}{2} \right)},
\fe
which yields, therefore, 
\ie
   \frac{\Gamma^{(\mathrm{KR})}}{\Gamma^{(\mathrm{GR})}}
   \;=\;
   \,e^{8 \pi  \omega  \left[(1-l)^{\Bar{\alpha} }-1\right] \left( M - \frac{\omega}{2} \right)} \approx 1 - 8\, l \, \pi  \Bar{\alpha}  \, \omega  \left( M - \frac{\omega}{2} \right).
\fe

\begin{figure}
    \centering
      \includegraphics[scale=0.45]{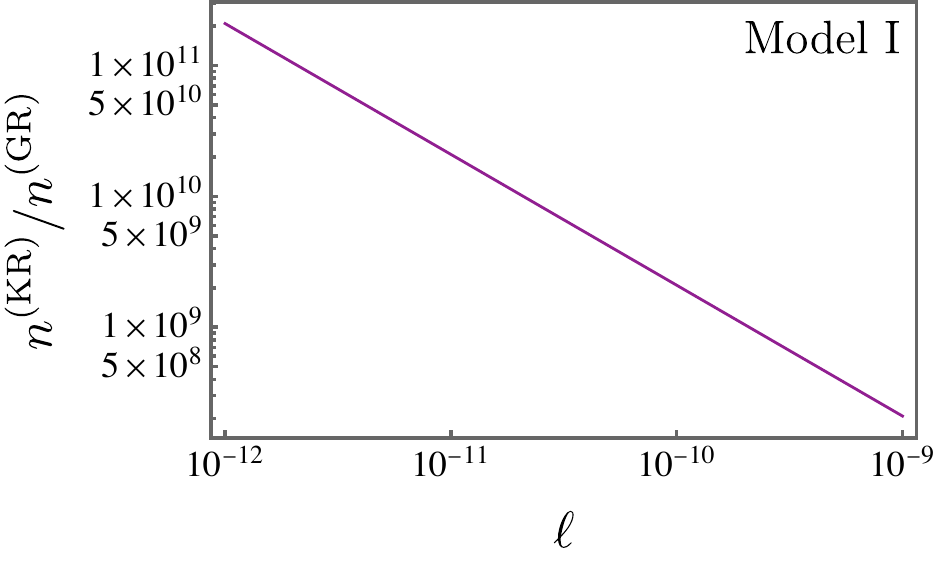}
       \includegraphics[scale=0.45]{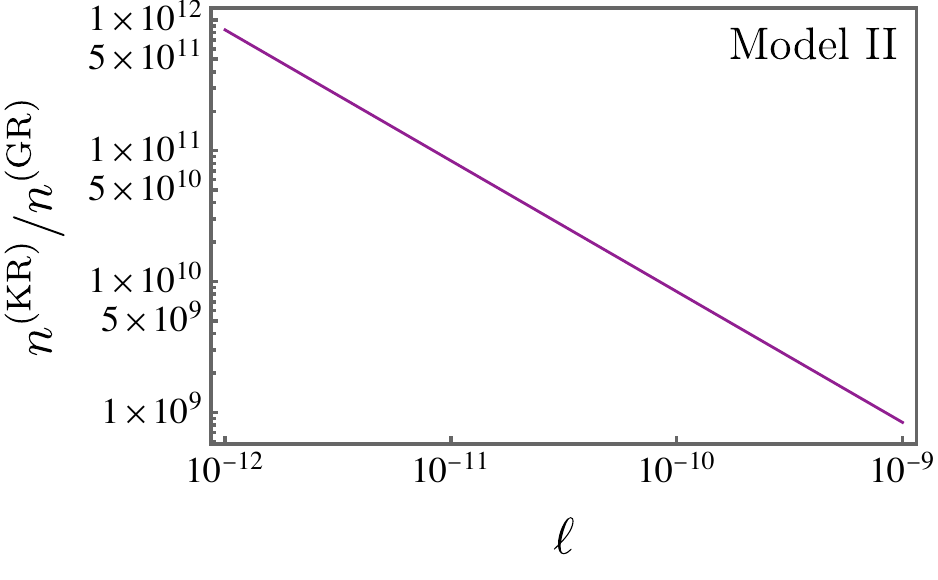}
    \caption{ The ratio $n^{(\mathrm{KR})}/n^{(\mathrm{GR})}$ as a function of $\ell$ is shown for both models, using a logarithmic scale on both axes.}
    \label{depaturefromgrparticle}
\end{figure}

Finally, in order to accomplish the deviation of general relativity on the particle creation, we write
\ie
\begin{split}
& \Bar{n} =  \frac{n^{(\mathrm{KR})}}{n^{(\mathrm{GR})}} =  \frac{\frac{\Gamma^{(\mathrm{KR})}}{\Gamma^{(\mathrm{GR})}}}{1 - \frac{\Gamma^{(\mathrm{KR})}}{\Gamma^{(\mathrm{GR})}}}  = \frac{1}{e^{-8 \pi  \omega  \left[(1-l)^{\Bar{\alpha} }-1\right] \left(M - \frac{\omega}{2} \right)}-1} \\
& \approx \frac{l \left(\Bar{\alpha} ^2 \left(16 \pi ^2 \omega ^2 (\omega -2 M)^2+1\right)-1\right)}{48 \pi  \Bar{\alpha}  \omega  (2 M-\omega )}+\frac{1}{l \left(8 \pi \Bar{\alpha}  M \omega -4 \pi \Bar{\alpha}  \omega ^2\right)}-\frac{-4 \pi \Bar{\alpha}  \omega ^2-\Bar{\alpha} +8 \pi  \Bar{\alpha}  M \omega +1}{16 \pi  \Bar{\alpha}  M \omega -8 \pi  \Bar{\alpha}  \omega ^2}.
\end{split}
\fe
In Fig. \ref{depaturefromgrparticle}, we display $\bar{n}$ as a function of $\ell$ for both Models studies in this paper.

In Tab. \ref{tab:evap-time}, we present the values of $\bar{t}$ alongside the bounds derived from Solar System tests reported in the literature \cite{yang2023static}. In general lines, since $|\ell| \le 10^{-10}$ (or even smaller), the largest deviations in evaporation time remain at the level of $\mathcal{O}(10^{-9})$. For Shapiro time delay ($|\ell| \approx 10^{-13}$), these deviations are even tinier, at the level of $10^{-12}\text{--}10^{-13}$. Negative $\ell$ ($\ell_{\min} < 0$), on the other hand, leads to an increase in the evaporation time ($+\,{\Bar{\beta}}\,|\ell|$), whereas positive $\ell$ reduces the evaporation time. Even for a solar--mass black hole (lifetime $\sim10^{67}$~years in GR), a correction at the level of $10^{-9}$ would shift the total evaporation time in a very small manner.

\begin{table}[h!]
\centering
\caption{The bounds for $\ell$ derived from Solar System tests \cite{yang2023static} are shown alongside $\bar{t}$.}
\label{tab:evap-time}
\begin{tabular}{lc | c | c}
\hline\hline
\multirow{2}{*}{\textbf{Constraint}} 
 & \multirow{2}{*}{\(\ell\)} 
 & \multicolumn{1}{c}{\textbf{Model I} } 
 & \multicolumn{1}{c}{\textbf{Model II} }
 \\
 & & \( \bar{t} \approx 1 - 5\,\ell\)
   & \( \bar{t} \approx 1 - 2\,\ell\) 
 \\
\hline
\textbf{Mercury precession} \\ \hline
\quad 
 & \(\,\ell_{\min}= -3.7\times10^{-12}\)
 & \( 1 +1.85\times10^{-11}\)
 & \( 1 +7.4\times10^{-12}\)
 \\
\quad
 & \( \ell_{\max} = +1.9\times10^{-11}\)
 & \(1 -9.5\times10^{-11}\)
 & \(1 -3.8\times10^{-11}\)
 \\
\hline
\textbf{Light deflection} \\ \hline
\quad 
 & \(\,\ell_{\min}= -1.1\times10^{-10}\)
 & \(1  +5.5\times10^{-10}\)
 & \(1 +2.2\times10^{-10}\)
 \\
\quad 
 & \(\ell_{\max}=+5.4\times10^{-10} \)
 & \(1 -2.7\times10^{-9}\)
 & \(1 -1.08\times10^{-9}\)
 \\
\hline
\textbf{Shapiro time delay} \\ \hline
\quad
 & \(\,\ell_{\min} = -6.1\times10^{-13}\)
 & \( 1 +3.05\times10^{-12}\)
 & \(1  +1.22\times10^{-12}\)
 \\
\quad 
 & \(\,\ell_{\max}= +2.8\times10^{-14}\)
 & \(1 -1.4\times10^{-13}\)
 & \(1 -5.6\times10^{-14}\)
 \\
\hline\hline
\end{tabular}
\end{table}


\section{Conclusion}

In this paper, we examined particle creation and the evaporation process within Kalb--Ramond gravity, employing two solutions from the literature, denoted here as Model I \cite{yang2023static} and Model II \cite{Liu:2024oas}.

Starting with Model I, we focused on bosonic modes, using the scalar field in curved spacetime to address \textit{Hawking} radiation. This allowed us to calculate the \textit{Hawking} temperature, $T$, using Bogoliubov coefficients, observing that increasing \(\ell\) led to a higher temperature. We then explored \textit{Hawking} radiation through the tunneling process, being based on the energy conservation. Our findings showed that the emission spectra deviated from the typical black body radiation due to Lorentz--violating corrections associated with \(\ell\). The resulting particle number density, \(n_1\), demonstrated that increasing \(\ell\) also increased particle density. Fermionic modes were subsequently considered, yielding the particle density \(n_{\psi1}\). Similar to the bosonic case, \(n_{\psi1}\) increased with \(\ell\).  Lastly, the evaporation process for Model I revealed that a larger Lorentz--violating parameter \(\ell\) shortened the black hole’s lifetime \(t_{\text{evap}}\), which was derived analytically.

For Model II, we followed a similar approach, analyzing both bosonic and fermionic modes along with the evaporation process. Here, the \textit{Hawking} temperature \(T_2\) was obtained through the Klein–Gordon equation in curved spacetime, again showing an increase in temperature with \(\ell\). The tunneling process also yielded particle densities \(n_2\) for bosons and \(n_{\psi2}\) for fermions, both of which increased as \(\ell\) grew. The evaporation process, characterized by the evaporation lifetime $t_{\text{evap2}}$, was found to decrease with increasing $\ell$, similar to the behavior observed in Model I.

Comparing the two models, particle densities \(n_1\) and \(n_2\) (bosonic) and \(n_{\psi1}\) and \(n_{\psi2}\) (fermionic) were found to be higher in Model I than in Model II, and both models produced higher densities than the Schwarzschild case. In terms of evaporation time, Model I had a shorter lifetime than Model II, as \(\ell \ll 0\). While both models evaporated faster than the Schwarzschild black hole, Model I consistently exhibited a more rapid evaporation rate than Model II.  Furthermore, in the analysis of greybody factors, Model II displayed higher magnitudes for both bosons and fermions compared to Model I and the Schwarzschild case. When comparing the spin of the particles, the greybody factors for fermions consistently exceeded those for bosons within each model. In contrast, Model I exhibited lower magnitudes for both particle types relative to the Schwarzschild case. Furthermore, we analyzed the deviation of our results from those predicted by general relativity.

As a future perspective, our investigation can be extended to include two new solutions within the framework of effective quantum gravity, as presented in Ref. \cite{zhang2024black}. This and other Lorentz--violating extensions are currently under development.


\section*{Acknowledgments}
\hspace{0.5cm}

A. A. Araújo Filho acknowledges support from the Conselho Nacional de Desenvolvimento Científico e Tecnológico (CNPq) and the Fundação de Apoio à Pesquisa do Estado da Paraíba (FAPESQ) under grant [150891/2023-7]. The author also extends gratitude to A. Övgün for his assistance with the regularization of divergent integrals in the calculations.

\section{Data Availability Statement}

Data Availability Statement: No Data associated in the manuscript


\bibliographystyle{ieeetr}
\bibliography{main}

\end{document}